\documentclass[prb,aps,twocolumn,superscriptaddress,showpacs,longbibliography]{revtex4-2}
\usepackage{natbib}
\usepackage{graphicx}
\usepackage{amssymb}
\usepackage{wrapfig}
\usepackage{epstopdf}
\usepackage{amsmath}
\usepackage{xspace} 
\usepackage{color}

\begin{document}
\title{BCS-BEC crossover of polaritonic condensates in mass-imbalanced semimetal/semiconductor microcavities}
\author{Thi-Hau Nguyen}
\affiliation{Hanoi University of Mining and Geology, 18 Pho Vien, Dong Ngac, Hanoi 10072, Vietnam;}
\affiliation{Graduate University of Science and Technology, Vietnam Academy of
Science and Technology, 18 Hoang Quoc Viet, Hanoi 10072, Vietnam.}
\author{Minh-Tien Tran}
\affiliation{Institute of Physics, Vietnam Academy of Science and Technology, 10 Dao Tan, Hanoi 10072, Vietnam;}
\author{Van-Nham Phan}
\thanks{{Corresponding author: phanvannham@duytan.edu.vn}}
\affiliation{Institute of Research and Development, Duy Tan University, 3 Quang Trung, Danang  550000, Vietnam}
\affiliation{Faculty of Natural Sciences, Duy Tan University, 3 Quang Trung, Danang 550000, Vietnam}

\begin{abstract}
The impacts of the mass imbalance and Coulomb interaction on the complex phase structures of the polaritonic condensates and their Bardeen-Cooper-Schrieffer (BCS)--Bose-Einstein condensation (BEC) crossover in semiconductor and semimetal microcavities are investigated. In the framework of the unrestricted Hartree-Fock approximation, a two-band electron-hole model involving photon mode is analyzed by treating Coulomb attraction and light-matter coupling on equal footing. The single-particle spectral functions and the luminescence properties are then examined. In the semiconducting regime, a positive band gap stabilizes tightly bound excitons and yields predominantly BEC-type excitoniclike polaritonic condensates at low density, while increasing excitation density and reducing mass imbalance drives a continuous crossover toward BCS-type pairing with intermediate and photoniclike polaritonic character. In contrast, the semimetallic regime favors itinerant electron-hole pairing, with BCS-type condensates dominating and BEC excitoniclike coherence emerging only at sufficiently strong Coulomb interaction and large mass imbalance situations. The evolution of luminescence spectra provides clear spectroscopic signatures of these crossover phenomena, offering a unified framework for understanding and controlling polaritonic condensates in microcavity systems.
\end{abstract}

\date{\today}
\maketitle

\section{Introduction}

Quantum coherent phases emerging from electron-hole systems have long been central topics in condensed matter physics, beginning with the prediction of the excitonic condensation state in semimetallic (SM) or narrow-gap semiconducting (SC) materials driven by electron-hole Coulomb attraction more than 60 years ago~\cite{Mo61,Kno63,JRK67,PRL.19.439,HR68b}. With the development of SC microcavities, it became possible to strongly couple excitons to quantized cavity modes, forming exciton-photon hybrid or polariton quasiparticles that combine matterlike with photonlike dispersions~\cite{RMP.82.1489,BKY14}. Because polaritons inherit an extremely small effective mass, typically on the order of $10^{-4}$ of the free-electron mass, they exhibit macroscopic quantum coherence at much higher temperatures than conventional atomic or excitonic systems~\cite{Iman2020}. This ultralight mass has attracted substantial interest because it permits Bose-Einstein condensation (BEC) even at room temperature in several SC microcavity platforms{, including organic, GaN, and ZnO systems~\cite{PRL.98.126405,APL.93.051102,NPho.4.371,RMP.85.299}}. As a result, microcavity polaritons have emerged as a versatile platform for studying light-matter many-body states and high-temperature quantum coherence, extending the landscape of accessible solid-state quantum phases~\cite{Iman2020}.

At low excitation densities, polaritonic condensation (PC) demonstrates signatures of BEC, including macroscopic ground-state occupation, long-range phase coherence, and nonlinear interaction-driven blueshifts~\cite{DWSBY02,Kas06,PRX.11.011018}. At high densities, Coulomb screening drives the system toward a Bardeen-Cooper-Schrieffer (BCS)--like regime of weakly correlated electron-hole Cooper pairs that can support photon lasing in case of sufficiently large density~\cite{PRB.76.201305,DWSBY02}. Between these limits resides the long-anticipated polaritonic BCS regime, in which electron-hole pairs retain significant Coulomb correlations despite their spatial overlap and continue to couple coherently to the confined cavity photon mode. Early theoretical studies predicted a continuous BEC-BCS crossover of the polaritonic condensate in microcavity systems~\cite{KO11,NJP.14.065001,Yamaguchi2015}, and recent experiments have observed fermionic gain and spectral signatures consistent with a BCS polaritonic phase~\cite{PRX.11.011018}. These developments indicate that microcavity polaritons can preserve a rich hierarchy of coherent states, ranging from strongly bound excitons to overlapping fermion pairs and ultimately photon-dominated condensates.

A quantitative understanding of this crossover requires a theoretical framework that incorporates Coulomb interactions and light-matter coupling on equal footing. Variational, renormalization, and Green’s function approaches have shown that the competition between excitonic binding and photonic hybridization governs the evolution of condensate order parameters, correlation functions, and luminescence spectra across the crossover~\cite{kamide2010,KO11,NJP.14.065001,Yamaguchi2015,PBF16,PhysB.573.72}. Complementary nonequilibrium analyses further highlight the interplay between coherent pairing, pumped reservoirs, and dissipation in shaping polaritonic phases~\cite{NJP.14.065001,BKY14,PRX.11.011018}. Despite these efforts, several crucial microscopic parameters remain insufficiently explored. One such parameter is the electron-hole mass imbalance, which strongly influences the nature of electron-hole pairing~\cite{EKOF14,PRB.109.085105,PRB.110.235143,PRB.111.245111}. In realistic semiconductor quantum wells and two-dimensional materials, the electron-hole mass ratio varies widely, from nearly unity in transition-metal dichalcogenide monolayers~\cite{2DM.2.022001,Jiwon2014} to well below $0.15$ in systems such as GaAs, GaN, and AlN~\cite{PhysicaB.210.1,SGJ1995,Suzuki1995}. Such variations lead to substantial changes in excitonic binding energies, pair sizes, and coherence lengths. Mass-imbalanced excitonic condensates have been investigated in both continuum and lattice-based models, revealing that mass asymmetry can substantially alter pairing strength, critical temperatures, and BEC-BCS crossover signatures~\cite{BF06,KEFO13,PRB.109.085105,PRB.110.235143}. In microcavities, recent theoretical analyses incorporating mass asymmetry have shown that the excitonic-like, intermediate, and photonic-like PC phases respond differently to changes in the mass ratio, reshaping phase boundaries and bound-state characteristics~\cite{PRB.104.245404,PRR.2.023089,PhDTh-Tiene,PRB.111.245111}. Nevertheless, a comprehensive microscopic description of how mass imbalance, Coulomb attraction, and excitation density together shape the BEC-BCS crossover of PCs remains incomplete.

In this work, we investigate the impacts of electron-hole mass asymmetry and Coulomb interactions on the BEC-BCS crossover of the complex structure of the PCs in equilibrium microcavities, focusing on two-band models relevant to optical SC and SM microcavities~\cite{kamide2010,KO11,NJP.14.065001,Yamaguchi2015,PBF16,PhysB.573.72,PRB.111.245111}. The Hamiltonian is analyzed within the unrestricted Hartree-Fock approximation (UHFA), an extension of the standard Hartree-Fock method that retains all off-diagonal expectation values and thereby accommodates spontaneous symmetry breaking in electron-hole and light-matter channels~\cite{SC08}. The UHFA has proven effective for elucidating excitonic condensation and phase diagrams in systems undergoing SM-SC transitions, including those with mass imbalance, as well as for modeling PCs in microcavities~\cite{PRB.107.115106,PRB.109.085105,PRB.110.235143,PRB.111.245111}. By treating electron-hole attraction and light-matter coupling on equal footing, we self-consistently determine the PC order parameter. Through detailed analysis of the interplay between excitonic and photonic components, we identify ground states that may be excitoniclike PC (EPC), intermediate PC (IPC), or photoniclike PC (PPC) states. Their corresponding BEC-BCS crossover behaviors are examined through the momentum distribution of the electron-hole pairing amplitude and the photonic density~\cite{kamide2010,KO11,NJP.14.065001,Yamaguchi2015}. Furthermore, we characterize luminescence and quasiparticle structure of the stabilities via the single-particle spectral functions of electrons, holes, together with the excitonic polarization and photon mode response functions~\cite{PRB.50.1119,PRB.81.081302,SSKSKSSNKF12,PBF16}. Our findings reveal how mass imbalance qualitatively reshapes the condensate landscape, redistributes coherence between matter and light sectors, and produces experimentally observable signatures in luminescence and spectral properties. Taken together, this work provides a unified microscopic picture of correlated PCs in mass-imbalanced SM and SC microcavities.

This paper is organized as follows. In Sec.~II, we introduce the equilibrium electron-hole-photon model in a microcavity and specify the Hamiltonian of the mass-imbalanced system. Sec.~III presents the unrestricted Hartree-Fock formalism and the self-consistent equations determining the condensate order parameters. In Sec.~IV, we derive the single-particle spectral functions and the luminescence responses of excitonic polarization and cavity photons in the present approach. The numerical results and their physical interpretation are discussed in Sec.~V, where we analyze the ground-state phase diagrams, BCS-BEC crossover behavior, and spectroscopic signatures of EPC, IPC, and PPC under the influence of mass imbalance and Coulomb interaction. Finally, Sec.~VI summarizes the main results and conclusions of this work.

\section{Hamiltonian}
Unlike closed systems in true thermal equilibrium, electron-hole-photon systems are intrinsically open, dissipative requiring external pumping to balance cavity losses. However, when the polariton thermalization time is shorter than their lifetime, a condition achievable in high-quality microcavities, the system can reach a quasithermal equilibrium state~\cite{RMP.82.1489,BKY14,Iman2020}. In this equilibrium limit, the ground state of the system and its dependence on excitation density become an interesting theoretical problem~\cite{NJP.14.065001,Yamaguchi2015}. In this study, we consider a 2D electron-hole-photon system in thermal equilibrium confined within a planar microcavity and the total many-particle Hamiltonian generally includes both noninteracting and interaction parts. In the momentum space, the noninteracting part can be written as
\begin{equation}\label{eq2}
\mathcal{H}_{0} = \sum_{\mathbf{k}}\left(\varepsilon^{e}_{\mathbf{k}} e^{\dagger}_{\mathbf{k}} e_{\mathbf{k}}+\varepsilon^{h}_{\mathbf{k}} h^{\dagger}_{\mathbf{k}} h_{\mathbf{k}}
\right)+\sum_{\mathbf{q}}\omega_{\mathbf{q}}\psi^{\dagger}_{\mathbf{q}}\psi_{\mathbf{q}},
\end{equation}
where $ e^{\dagger}_{\mathbf{k}} (e_{\mathbf{k}}) $, $ h^{\dagger}_{\mathbf{k}} (h_{\mathbf{k}}) $, and $ \psi^{\dagger}_{\mathbf{k}} (\psi_{\mathbf{k}}) $ are the creation (annihilation) operators of the spinless conduction electrons, valence holes, and photons with momentum $\mathbf{k}$, respectively, corresponding to their dispersion relations $\varepsilon^{e}_{\mathbf{k}}, \varepsilon^{h}_{\mathbf{k}}$ and $\omega_{\mathbf{k}}$. In the tight-binding approximation, the dispersion relation of the electrons/holes reads
\begin{equation}\label{eq3}
\varepsilon_{\mathbf{k}}^{e/h} = - 2t^{e/h}(\cos k_x + \cos k_y)+\frac{E_g + 8t^{e/h} - \mu}{2}.
\end{equation}
Here $t^{e/h}$ is the nearest-neighbor hopping amplitude of electrons/holes in lattice, which is inversely proportional to their corresonding effective mass. {Near $\mathbf{k}=0$ these bands are equivalent to isotropic parabolic bands. However, unlike a pure parabola the cosine form naturally has finite bandwidth and van Hove singularities at the Brillouin-zone edges. Such non-parabolicity can be important when the Fermi level or exciton center-of-mass momentum is large. We emphasize that the square lattice is not meant to represent any specific crystal, it is a generic model that captures band curvature and finite bandwidth effects.} In Eq.~\eqref{eq3}, $E_g$ represents the energy gap between the bottom of the conduction band and the top of the valence band, thereby distinguishing between SC, $E_g > 0$ and SM, $E_g<0$, states. The dispersion relation of photons $\omega_{\mathbf{q}}$ in microcavity is
\begin{equation}\label{eq5}
\omega_{\mathbf{q}} = \sqrt{(c \mathbf{q})^2 + \omega_c^2} - \mu,
\end{equation}
where $c$ is the speed of light in the microcavity and in the nature unit one sets $c=1$. In Eq.~\eqref{eq5}, $\omega_{c}$ is a zero-point cavity frequency. The matter-light correlation depends on the overlap between the photonic band and those of electron-hole bands. Therefore, to address the bound state formed by an electron-hole pair and photons inside the microcavity, one introduces the detuning parameter $d = \omega_c - E_g$~\cite{kamide2010,KO11}. When $d$ is small, the excitation energy of the electron-hole pair lies close to the bottom of the photon band, resulting in an enhanced photonic contribution to the PC. In contrast, for a large detuning, the overlap between the photon and the electron-hole bands becomes weak, and consequently, the photonic contribution to the PC becomes less significant ~\cite{KO11,NHAM2017}. {The quantity $\mu$ in Eqs.~\eqref{eq3} and \eqref{eq5} denotes the chemical potential, which is adjusted to fix the total excitation density defined as the sum of electron-hole pairs and cavity photons,
\begin{equation}\label{eq4}
n = \frac{1}{N}\sum_{\mathbf{q}} \langle \psi^{\dagger}_{\mathbf{q}}\psi_{\mathbf{q}} \rangle 
+ \frac{1}{2N}\sum_{\mathbf{k}}\left(\langle e^{\dagger}_{\mathbf{k}} e_{\mathbf{k}} \rangle + \langle h^{\dagger}_{\mathbf{k}} h_{\mathbf{k}} \rangle \right),
\end{equation}
where $N$ is the number of lattice sites. Here the factor $1/2$ in the second term of Eq.~\eqref{eq4} ensures that electrons and holes are counted as pairs. In the present formulation, cavity photons are strongly coupled to electron-hole pairs via the light-matter interaction, leading to hybridized polariton modes. As a result, the system can be viewed as a single interacting quasiparticle ensemble in which the conserved quantity is the total number of excitations. The chemical potential therefore controls the total excitation density of the system, and it naturally appears the same in the dispersions of electrons, holes, and photons within the grand-canonical formalism~\cite{kamide2010,KO11}.}

In the electron-hole-photon system, the interacting part of the Hamiltonian includes the electron-hole Coulomb attraction and the matter-light coupling. In the momentum space, one delivers
\begin{align}\label{eq6}
\mathcal{H}_\textrm{int}=&- \frac{U}{N} \sum_{\mathbf{k}\mathbf{k}'\mathbf{q}} 
e_{\mathbf{k}+\mathbf{q}}^{\dagger} e^{}_{\mathbf{k}} h_{\mathbf{k}'-\mathbf{q}}^{\dagger} h^{}_{\mathbf{k}'}\nonumber\\
&- \frac{g}{\sqrt{N}} \sum_{\mathbf{k}\mathbf{q}} (e_{\mathbf{k}+\mathbf{q}}^\dagger h_{-\mathbf{k}}^\dagger \psi^{}_{\mathbf{q}} + \textrm{H.c.}),
\end{align}

In this expression of the interaction Hamiltonian, the first term addresses the Coulomb attraction between conduction-band electrons and valence-band holes. {For simplicity, we approximate the electron-hole Coulomb interaction by a momentum-independent coupling constant $U$. This corresponds to assuming that the Coulomb interaction is strongly screened so that it can be approximated by an effective short-range interaction in real space. Such a Falicov-Kimball interaction approximation is widely used in minimal models of excitonic condensation, where the essential physics is the formation of electron-hole pairs, while the detailed momentum dependence of the Coulomb potential is neglected~\cite{PBF10,IPBBF08,SEO11,ZIBF12,EKOF14}}. The second term expresses the Dicke interaction, which was originally introduced to describe the coherent coupling between an ensemble of two-level atoms and a single-mode electromagnetic field~\cite{dick1954} and has been widely used to examine the polariton condensate in optical microcavity~\cite{PRB.64.235101,PRL.96.230602,RMP.82.1489,RMP.85.299}. In the present work, we incorporate both the Coulomb attraction from the Falicov-Kimball scenario and the matter-light coupling of the Dicke picture on an equal footing within a single compact Hamiltonian~\cite{NJP.14.065001,PBF16,PRB.111.245111}. Such a treatment enables us to capture the intrinsic BCS-BEC characteristics of PCs in a microcavity.

\section{UHFA}

To analyze the ground-state properties of the interacting electron-hole-photon system, we treat the total Hamiltonian $\mathcal{H}=\mathcal{H}_0+\mathcal{H}_\textrm{int}$ within the UHFA. This mean-field approach allows for the spontaneous breaking of $U(1)$ symmetry associated with the formation of coherent states, while the Coulomb attraction and light-matter coupling are handled on an equal footing. To proceed, the interaction terms in $\mathcal{H}_{\mathrm{int}}$ in Eq.~\eqref{eq6} are decoupled by expanding the field operators $\hat{O}$ around their thermodynamic expectation values $\langle \hat{O} \rangle$, i.e., $\hat{O} = \langle \hat{O} \rangle + \delta \hat{O}$, and neglecting terms of second order in the fluctuations $\delta \hat{O}$. For the quartic Coulomb interaction, by retaining the distinct density and pairing channels, we have
\begin{align}
e_{\mathbf{k}+\mathbf{q}}^{\dagger}&e^{}_{\mathbf{k}}h_{\mathbf{k}'-\mathbf{q}}^{\dagger} h^{}_{\mathbf{k}'} \approx \, \delta_{\mathbf{q},0} ( n^e_{\mathbf{k}} h^{\dagger}_{\mathbf{k}'} h^{}_{\mathbf{k}'} + n^h_{\mathbf{k}'} e^{\dagger}_{\mathbf{k}} e^{}_{\mathbf{k}} - n^e_{\mathbf{k}} n^h_{\mathbf{k}'} ) \nonumber \\
&+ \delta_{\mathbf{k},-\mathbf{k}'} ( d_{\mathbf{k}} e^{\dagger}_{\mathbf{k}+\mathbf{q}} h^{\dagger}_{-\mathbf{k}-\mathbf{q}} + d^\ast_{\mathbf{k}+\mathbf{q}} h^{}_{-\mathbf{k}} e^{}_{\mathbf{k}} - |d^{}_{\mathbf{k}}|^2),
\end{align}
where $n^{e(h)}_{\mathbf{k}} = \langle e^{\dagger}_{\mathbf{k}} e^{}_{\mathbf{k}} \rangle$ ($ \langle h^{\dagger}_{\mathbf{k}} h^{}_{\mathbf{k}} \rangle$) is the electron (hole) density distribution function and $d_{\mathbf{k}} = \langle e_{\mathbf{k}}^{\dagger} h_{-\mathbf{k}}^{\dagger} \rangle$ is the electron-hole pairing amplitude. Similarly, the light-matter interaction is linearized by treating the photon field as a macroscopic classical field $\psi_{\mathbf{q}} \to \langle \psi_{\mathbf{q}} \rangle + \delta \psi_{\mathbf{q}}$, where $\langle \psi_{\mathbf{q}} \rangle$ characterizes the photonic condensate order parameter, so
\begin{align}
e_{\mathbf{k}+\mathbf{q}}^\dagger h_{-\mathbf{k}}^\dagger \psi^{}_{\mathbf{q}} \approx \delta_{{\bf q},0} (d_{\mathbf{k}}\psi^{}_{\mathbf{q}}+\langle  \psi^{}_{\mathbf{q}}\rangle e_{\mathbf{k}}^\dagger h_{-\mathbf{k}}^\dagger),
\end{align}

Under these approximations, the total mean-field Hamiltonian can be expressed compactly. Using the Nambu basis $\Psi^\dagger_{\mathbf{k}} = (e^\dagger_{\mathbf{k}}, h^{}_{-\mathbf{k}})$, one can find its electron-hole part taking in the matrix form 
\begin{equation}
\mathcal{H}_\textrm{UHF}^\textrm{e-h} = \sum_{\mathbf{k}} \Psi_{\mathbf{k}}^\dagger \begin{pmatrix} \bar{\varepsilon}^e_{\mathbf{k}} & \Delta \\ \Delta^* & -\bar{\varepsilon}^h_{\mathbf{k}} \end{pmatrix} \Psi_{\mathbf{k}} + \mathcal{C},  
\label{eq7}
\end{equation}
where $\mathcal{C}$ contains constant energy shifts that would be neglected for the further evaluation. The diagonal elements in the $2\times 2$ matrix represent the renormalized dispersions 
\begin{equation}
\bar{\varepsilon}^{e(h)}_{\mathbf{k}} = \varepsilon^{e(h)}_{\mathbf{k}} - Un^{h(e)}, 
\label{eq8}
\end{equation}
with $n^{e(h)}=\sum_{\mathbf{k}}n^{e(h)}_{\mathbf{k}}/N$, which account for the renormalization of the band structure due to the attractive Hartree potential exerted by the opposite carrier type. The off-diagonal elements in the matrix in Eq.~\eqref{eq7} define the total PC order parameter
\begin{align}
\Delta &= -\frac{U}{N}\sum_{\mathbf{k}} d_{\mathbf{k}} - \frac{g}{\sqrt{N}} \langle \psi_0 \rangle \nonumber\\
&\equiv \Delta_\textrm{eh} + \Delta_\textrm{ph}.  
\label{eq9}
\end{align}

The gap function $\Delta$ is the central result of the UHFA analysis, demonstrating that the symmetry breaking field is a coherent superposition of the matter excitonic field, the so-called excitonic condensate order parameter $\Delta_\textrm{eh}$, and the cavity photon field, the so-called photonic condensate order parameter $\Delta_\textrm{ph}$. The interplay between $\Delta_\textrm{eh}$ and $\Delta_\textrm{ph}$ can characterize the complex phase structure of the cavity PC. In the cooperative regime $0 <\Delta_\textrm{eh}/\Delta_\textrm{ph} < 1$), the system resides in the PC state where the nonzero expectation value $\langle \psi_0 \rangle$ acts as an external field for the excitonic coherence and vice versa, the nonzero of $d_{\mathbf{k}}$ effectively raises the photonic coherent state.

The photonic part of the total mean-field Hamiltonian takes the form of a driven harmonic oscillator
\begin{equation}
\mathcal{H}^{\mathrm{ph}}_{\mathrm{UHF}} = \sum_{\mathbf{q}}\omega_\mathbf{q} \psi_\mathbf{q}^{\dagger} \psi_\mathbf{q} + \sqrt{N}(\Gamma \psi_0^{\dagger} + \Gamma^* \psi_0).
\end{equation}
The driving force $\Gamma = -g\sum_{\mathbf{k}} d_{\mathbf{k}}/N$ physically represents the macroscopic interband polarization density of the electron-hole system. In this mean-field picture, the collective coherence of the matter field $d_{\mathbf{k}}$ acts as a classical source current that displaces the photon vacuum. This establishes a crucial feedback loop that the electronic polarization generates a coherent photon field via $\Gamma$, which in turn enhances the pairing potential $\Delta$ via the Rabi coupling $g$, thereby reinforcing the original polarization.

To obtain the self-consistent equations, we perform the unitary Bogoliubov transformation by defining new quasiparticle fermionic operators
\begin{align}
\gamma_{+\mathbf{k}} &= u_{\mathbf{k}} e_{\mathbf{k}} + v_{\mathbf{k}} h_{-\mathbf{k}}^\dagger, \nonumber\\
\gamma_{-\mathbf{k}} &= -v_{\mathbf{k}} e_{\mathbf{k}} + u_{\mathbf{k}} h_{-\mathbf{k}}^\dagger,
\label{eq10}
\end{align}
where the coherence factors satisfy $u_{\mathbf{k}}^2 + v_{\mathbf{k}}^2 = 1$ and are explicitly derived as 
\begin{align}
v_{\mathbf{k}}^2 = \frac{1}{2} \left( 1 - \xi_{\mathbf{k}}/W_{\mathbf{k}} \right),\quad
u_{\mathbf{k}}^2 = \frac{1}{2} \left( 1 + \xi_{\mathbf{k}}/W_{\mathbf{k}} \right),  
\label{eq11}
\end{align}
with $\xi_{\mathbf{k}} = (\bar{\varepsilon}^e_{\mathbf{k}} + \bar{\varepsilon}^h_{\mathbf{k}})/2$ and $W_{\mathbf{k}} = \sqrt{\xi_{\mathbf{k}}^2 + |\Delta|^2}$. 

In the new fermionic representation, the electron-hole Hamiltonian in Eq.~\eqref{eq7} becomes diagonalized and reads
\begin{equation}
\mathcal{H}^{\textrm{e-h}}_{\textrm{dia}} = \sum_{\alpha=\pm,\mathbf{k}}E_{\mathbf{k}}^\alpha \gamma_{\alpha\mathbf{k}}^\dagger \gamma^{}_{\alpha\mathbf{k}},  
\end{equation}
with quasiparticle energies given by
\begin{equation}
E_{\mathbf{k}}^{\pm} = \frac{\bar{\varepsilon}^e_{\mathbf{k}} - \bar{\varepsilon}^h_{\mathbf{k}}}{2} \pm W_{\mathbf{k}}.  
\label{eq12}
\end{equation}

The self-consistent equations are determined by calculating the thermodynamic expectation values of the original operators using the inverse Bogoliubov transformation. The carrier densities and pairing amplitude are given by
\begin{align}
n_{\mathbf{k}}^e & = u_{\mathbf{k}}^2 f(E_{\mathbf{k}}^+) + v_{\mathbf{k}}^2 f(E_{\mathbf{k}}^-),\nonumber \\
n_{\mathbf{k}}^h & = 1 - v_{\mathbf{k}}^2 f(E_{\mathbf{k}}^+) - u_{\mathbf{k}}^2f(E_{\mathbf{k}}^-), \label{eq13}\\
d_{\mathbf{k}} & = u_{\mathbf{k}} v_{\mathbf{k}} \left[ f(E_{\mathbf{k}}^+) - f(E_{\mathbf{k}}^-) \right],\nonumber
\end{align}
where $f(E) = 1/(e^{\beta E} + 1)$ is the Fermi-Dirac distribution function with $\beta=1/T$ and $T$ is temperature.

To determine the macroscopic photon amplitude $\langle \psi_0 \rangle$, we invoke the Heisenberg equation of motion for the annihilation operator $i\hbar \partial_t\psi_0 = [\psi_0, \mathcal{H}]$. In the thermodynamic equilibrium ground state, the coherent field amplitude is stationary, implying the time derivative of the expectation value vanishes, i.e., $\partial_t\langle \psi_0 \rangle = 0$. Evaluating the commutator with the mean-field Hamiltonian yields the relation $\hbar\omega_0 \langle \psi_0 \rangle + \sqrt{N}\Gamma = 0$. This linear algebraic equation confirms that the cavity field is statically displaced from the vacuum by the collective matter polarization, resulting in the relation 
\begin{equation}
\langle \psi_0 \rangle = - \frac{\sqrt{N}\Gamma}{\hbar\omega_0}.  
\label{eq14}
\end{equation}

From the expressions for the expectation values Eqs.~(\ref{eq13}) and (\ref{eq14}), combined with the renormalized energy bands in Eq.~\eqref{eq8} and PC order parameter in Eq.~\eqref{eq9}, we establish a closed set of self-consistent equations. By solving this system numerically for a fixed total excitation density, we can extract the spectral properties and thermodynamic order parameters that characterize the ground state of the mass-imbalanced electron-hole-photon system.

\section{Single-particle spectral and luminescence functions}
To further characterize the condensation states and directly probe the gap opening in the quasiparticle excitation spectrum, we calculate the wave vector--resolved single particle spectral functions. These quantities are directly experimentally accessible via angle-resolved photoemission spectroscopy. The spectral function for the conduction electrons is defined as the imaginary part of the retarded Green's function
\begin{equation}
A^e(\mathbf{k}, \omega) = -\frac{1}{\pi} \mathrm{Im} \, G^e(\mathbf{k}, \omega).
\label{eq15a}
\end{equation}
Here, the frequency-dependence of the Green's function $G^e(\mathbf{k}, \omega)$ is related to its time counterpart $G^e(\mathbf{k}, t) = -i \theta(t) \langle \{ e_{\mathbf{k}}(t), e_{\mathbf{k}}^{\dagger}(0) \} \rangle$ via the Fourier transform
\begin{equation}
G^e(\mathbf{k}, \omega) = \int_{-\infty}^{\infty} dt \, e^{i(\omega + i\eta)t} G^e(\mathbf{k}, t),
\label{eq15b}
\end{equation}
where $\eta \to 0^+$ ensures convergence.

To evaluate the Green's function, we express the electron operators in terms of the diagonalized Bogoliubov quasiparticles $\gamma_{\pm \mathbf{k}}$ using the inverse of Eq.~\eqref{eq10} and in the diagonal basis, the time evolution of the quasiparticle operators is trivial $\gamma_{\pm \mathbf{k}}(t) = e^{-i E_{\mathbf{k}}^{\pm} t} \gamma_{\pm \mathbf{k}}(0)$. In this schedule, one finds an expression of the frequency-dependence of the Green's function for electrons
\begin{equation}
G^e(\mathbf{k}, \omega) = \frac{u_{\mathbf{k}}^2}{\omega - E_{\mathbf{k}}^+ + i\eta} + \frac{v_{\mathbf{k}}^2}{\omega - E_{\mathbf{k}}^- + i\eta},
\end{equation}
where the coherence factors $u_{\mathbf{k}}$ and $v_{\mathbf{k}}$ are given in Eqs.~\eqref{eq11} and the quasi-particle energies $E_{\pm}(\mathbf{k})$ are given in Eq.~\eqref{eq12}. Taking the imaginary part leads to the analytical expression for the electron spectral function
\begin{equation}
A^e(\mathbf{k}, \omega) = u_{\mathbf{k}}^2 \delta(\omega - E_{\mathbf{k}}^+) + v_{\mathbf{k}}^2 \delta(\omega - E_{\mathbf{k}}^-).
\label{eq15}
\end{equation}

The expression of Eq.~(\ref{eq15}) elucidates that in the condensed phase ($\Delta \neq 0$), the spectral weight of the electron is split between the two quasiparticle branches. The upper branch $E_{\mathbf{k}}^+$ carries weight $u_{\mathbf{k}}^2$, while the lower branch $E_{\mathbf{k}}^-$ acquires weight $v_{\mathbf{k}}^2$ due to the particle-hole mixing.

Similarly, for the valence holes, the spectral function is defined via the Green's function of the holes $G^h(\mathbf{k}, t) = -i \theta(t) \langle \{ h_{\mathbf{k}}(t), h_{\mathbf{k}}^{\dagger}(0) \} \rangle$. Using the reverse relation $h_{-\mathbf{k}}^{\dagger} = v_{\mathbf{k}} \gamma_{+\mathbf{k}} + u_{\mathbf{k}} \gamma_{-\mathbf{k}}$ derived from Eq.~\eqref{eq10} and noting that hole creation corresponds to electron annihilation in the valence band, the time evolution leads to poles at $-\omega = E_{\mathbf{k}}^{\pm}$. The resulting spectral function is
\begin{equation}
A^h(\mathbf{k}, \omega) = v_{\mathbf{k}}^2 \delta(\omega + E_{\mathbf{k}}^+) + u_{\mathbf{k}}^2 \delta(\omega + E_{\mathbf{k}}^-).
\end{equation}
This expression describes the probability of removing a hole (or adding an electron to the valence band) with energy $\omega$.

The optical properties of the system are characterized by the luminescence spectrum, which provides direct insight into the collective excitations of the condensates. Following the linear response theory framework, the luminescence spectrum of the PCs in microcavity is analyzed in the signatures of the response functions for the excitonic polarization and the cavity photon mode. {Similar to Eqs.~\eqref{eq15a} and \eqref{eq15b}, the luminescence spectrum for excitonic polarization is specified as 
\begin{equation}
A^\textrm{ex}(\mathbf{k}, \omega) = -\frac{1}{\pi} \mathrm{Im} G^\textrm{ex}(\mathbf{k},\omega),
\end{equation}
where 
\begin{align}
G^\textrm{ex}(\mathbf{k},\omega) &=-i\int_{-\infty}^{\infty} dt\, e^{i(\omega+i\eta)t} \theta(t)\langle [b_\mathbf{k}(t),b_\mathbf{k}^{\dagger}(0)]\rangle\nonumber\\
&\equiv -i\int_{0}^{\infty} dt\, e^{i(\omega+i\eta)t}\langle [b_\mathbf{k}(t),b_\mathbf{k}^{\dagger}(0)] \rangle,
\label{eq16a}
\end{align}
is the retarded excitonic Green's function with the exciton creation operator $b_{\mathbf{k}}^\dagger$ defined as
\begin{equation}
b_{\mathbf{k}}^\dagger =\frac{1}{\sqrt{N}} \sum_{\mathbf{p}} e_{\mathbf{k}+\mathbf{p}}^\dagger h_{-\mathbf{p}}^\dagger.
\end{equation}
Using the identity $\mathrm{Im}F=(F-F^{\ast})/(2i)$ for any complex function $F$, the spectral function can be expressed as the Fourier transform
of the commutator over the entire time axis~\cite{PRB.50.1119,PRB.81.081302,SSKSKSSNKF12,PBF16},}
\begin{equation}
A^\textrm{ex}(\mathbf{k}, \omega) = \frac{1}{2\pi} \int_{-\infty}^{\infty} dt \, e^{i\omega t} \langle [b^{}_{\mathbf{k}}(t), b_{\mathbf{k}}^\dagger(0)] \rangle,
\end{equation}
Similar to evaluating the single particle spectral functions above, we arrive at the final analytical expression for the excitonic response function
\begin{align}
A^{\textrm{ex}}&(\mathbf{k}, \omega) \nonumber\\
=& \frac{1}{N}\sum_{\mathbf{p}}\Big[|u_{\mathbf{k}+\mathbf{p}} v_{\mathbf{p}}|^2 \mathcal{F}(E_{\mathbf{p}}^+, E_{\mathbf{k}+\mathbf{p}}^+) \delta(\omega - E_{\mathbf{k}+\mathbf{p}}^+ + E_{\mathbf{p}}^+) \nonumber \\
&\quad\quad+|v_{\mathbf{k}+\mathbf{p}} v_{\mathbf{p}}|^2 \mathcal{F}(E_{\mathbf{p}}^+, E_{\mathbf{k}+\mathbf{p}}^-) \delta(\omega - E_{\mathbf{k}+\mathbf{p}}^- + E_{\mathbf{p}}^+) \nonumber \\
&\quad\quad+|u_{\mathbf{k}+\mathbf{p}} u_{\mathbf{p}}|^2 \mathcal{F}(E_{\mathbf{p}}^-, E_{\mathbf{k}+\mathbf{p}}^+) \delta(\omega - E_{\mathbf{k}+\mathbf{p}}^+ + E_{\mathbf{p}}^-) \nonumber \\
&\quad\quad+|v_{\mathbf{k}+\mathbf{p}} u_{\mathbf{p}}|^2 \mathcal{F}(E_{\mathbf{p}}^-, E_{\mathbf{k}+\mathbf{p}}^-) \delta(\omega - E_{\mathbf{k}+\mathbf{p}}^- + E_{\mathbf{p}}^-)\Big],
\label{eq16}
\end{align}
where the thermal occupation factor is defined as $\mathcal{F}(E_1, E_2) = f(E_1) - f(E_2)$ with $f(E)$ is the Fermi-Dirac distribution function. This expression describes the spectral weight of electron-hole recombination processes mediated by the interband transitions between the renormalized quasiparticle bands.

The response function of the cavity photon mode, often referred to as the luminescence function, is defined as
\begin{equation}
B^\textrm{ph}(\mathbf{q}, \omega) = \frac{1}{2\pi} \int_{-\infty}^{\infty} dt \, e^{i\omega t} \langle [\psi_{\mathbf{q}}(t), \psi_{\mathbf{q}}^\dagger(0)] \rangle,
\end{equation}
and in the UHFA, the response function is given by
\begin{equation}
B^\textrm{ph}(\mathbf{q}, \omega) = \delta(\omega - \omega_{\mathbf{q}}).
\label{eq17}
\end{equation}
The photonic spectral weight is, therefore, concentrated entirely at the cavity dispersion energy $\omega_{\mathbf{q}}$.

\section{Numerical results and discussion}

In this section, we present the numerical results obtained from the self-consistent solution of the UHFA equations derived in the previous section. Our primary objective is to elucidate the ground-state phase diagram of the PCs in SM and SC microcavities, with a specific focus on the interplay between the electron-hole mass imbalance and the Coulomb interaction. The numerical calculations are performed on a 2D square lattice with $N = 200 \times 200$ sites. To set the energy scale, we fix the nearest-neighbor hopping amplitude of the conduction electrons to $t^e = 1$. The mass imbalance between the charge carriers is controlled by the hole hopping amplitude $t^h$, which is varied in the range $0 < t^h \leq 1$. Since the effective mass is inversely proportional to the hopping integral, the condition $t^h \leq t^e$ implies that the valence holes are heavier than the conduction electrons, $m_h \geq m_e$. The limit $t^h = 1$ corresponds to the mass-symmetric case $m_h = m_e$, a regime that has been extensively studied in the context of electron-hole bilayers and standard excitonic condensation states~\cite{NHAM2017,NHAM2016,PhysB.573.72}. Conversely, the limit $t^h \to 0$ represents the extreme mass-imbalance scenario where holes become localized, corresponding to a flat valence band with infinite effective mass. As a typical situation, the cavity photon frequency and the light-matter coupling strength are fixed at $\omega_c = 0.5$ and $g = 0.2$, respectively. The system is studied in the grand canonical ensemble, where the chemical potential $\mu$ is tuned to control the total excitation density $n$, as defined in Eq.~\eqref{eq4}. By systematically varying the mass imbalance parameter $t^h$ and the Coulomb interaction strength $U$, we examine the stability regions of the EPC, PPC, and IPC and their corresponding BCS-BEC crossover. The BCS-BEC crossovers are inspected in two distinct regimes, the SC state, characterized by a negative detuning parameter $d = -0.5$, and the SM state, characterized by a positive detuning $d = 1$. This allows us to contrast the condensate properties across the SM-SC transition driven by the cavity detuning parameter.

\subsection{Semiconductor side}

\begin{figure}[b]
\includegraphics[width=0.47\textwidth]{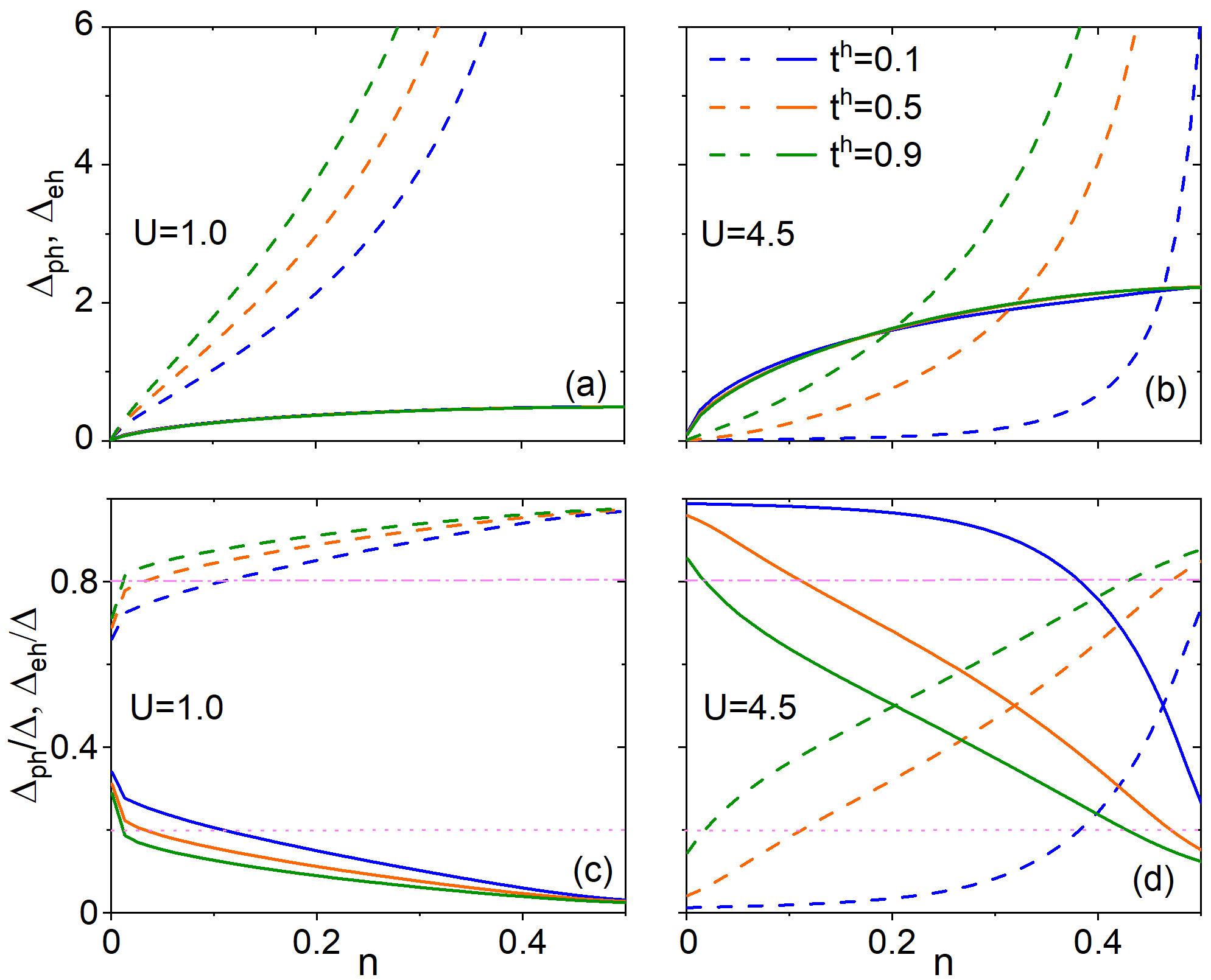}
\caption{Condensate order parameters as functions of the excitation density $n$ for different hole hopping amplitudes $t^h$ at detuning parameter $d=-0.5$. The top row shows the excitonic $\Delta_{\mathrm{eh}}$ (solid lines) and the photonic $\Delta_{\mathrm{ph}}$ (dashed lines), while the bottom row displays their relative weights in the condensates for Coulomb interaction (a) and (c) $U=1$  and (b) and (d) $U=4.5$. The horizontal dotted and dashed-dotted lines mark the fractions of 20$\%$ and 80$\%$, respectively.}
\label{f1}
\end{figure}

We begin our analysis by examining the stability and nature of the condensed phases in the SC regime with the detuning $d = -0.5$ corresponding to a positive band gap $E_g = 1$. Figure~\ref{f1}(a) and \ref{f1}(b) display the evolution of the excitonic $\Delta_{\mathrm{eh}}$ and photonic $\Delta_{\mathrm{ph}}$ condensate order parameters as functions of the total excitation density $n$ for different values of the hole transfer integral $t^h$ and two typical values of weak and strong Coulomb interaction strengths $U$ [$U=1$ in panel (a) and $U=4.5$ in panel (b)]. A universal feature observed in both interaction regimes is the monotonic increase of the order parameters with excitation density. However, the growth rates of the two components differ significantly. The photonic order parameter $\Delta_{\mathrm{ph}}$ exhibits a rapid increase with $n$. This behavior is characteristic of the Bose-like nature of the cavity photons that as the chemical potential approaches the bottom of the photon band, the macroscopic occupation of the zero-momentum photon mode increases dramatically, driving the order parameter. In contrast, the excitonic order parameter $\Delta_{\mathrm{eh}}$ tends to saturate at higher densities. This saturation is a direct consequence of the composite fermionic nature of the excitons. As the density increases, the available phase space for electron-hole pairing is restricted due to the Pauli exclusion principle. Consequently, the efficiency of pairing diminishes, preventing the unbounded growth of the excitonic amplitude even as the total particle number rises.

In the weak coupling regime, the photonic order parameter $\Delta_{\mathrm{ph}}$ dominates over the excitonic component $\Delta_{\mathrm{eh}}$ across the entire density range [see Fig.~\ref{f1}(a)]. This indicates that the condensate is primarily photonic in nature or a PPC. As the excitation density increases, the chemical potential rises rapidly towards the photon band edge, favoring the macroscopic occupation of the photon mode. The excitonic component remains small, reflecting the weak binding of electron-hole pairs. Conversely, in the strong coupling regime, $\Delta_{\mathrm{eh}}$ becomes significant, comparable with or even larger than $\Delta_{\mathrm{ph}}$ at low densities [see Fig.~\ref{f1}(b)]. This signifies the formation of a robust EPC. However, as the density increases, $\Delta_{\mathrm{eh}}$ saturates due to phase-space filling effects, while $\Delta_{\mathrm{ph}}$ continues to grow, eventually dominating at high densities. This crossover highlights the transition from EPC to PPC. 

The effect of mass imbalance is also evident. Increasing the hole hopping integral $t^h$ or reducing mass imbalance enhances the photonic order parameter $\Delta_{\mathrm{ph}}$, particularly at higher densities. This is because a larger $t^h$ corresponds to a wider valence band and a lower density of states (DOS), causing the chemical potential to rise more rapidly and reach the photon band edge at lower excitation densities. The excitonic order parameter $\Delta_{\mathrm{eh}}$, on the other hand, is almost insensitive to variations in $t^{h}$, especially for weak Coulomb interaction, indicating that excitonic pairing is faintly governed by the degree of mass imbalance. This weak dependence reflects the fact that, in this regime, electron-hole pairing is weak and broadly distributed in momentum space, so that modifying the hole dispersion only slightly affects the total pairing amplitude. At stronger Coulomb interaction, the overall magnitude of $\Delta_{\mathrm{eh}}$ is enhanced and a more visible dependence on $t^{h}$ emerges, as increased hole mobility improves the phase-space overlap between electron and hole states. Nevertheless, even in this case the variation of $\Delta_{\mathrm{eh}}$ with mass imbalance remains modest, leading to the robustness of excitonic binding once Coulomb attraction dominates over kinetic-energy effects.

To quantitatively clarify the competition between the excitonic and photonic components and to map the distinct condensate phases, we address in Figs.~\ref{f1}(c) and \ref{f1}(d) the evolution of the fractional order parameters $\Delta_{\mathrm{ph}}/\Delta$ and  $\Delta_{\mathrm{eh}}/\Delta$ as functions of the excitation density $n$. Based on the relative weights of these components, we classify the condensate nature into three distinct regimes with EPC when the photonic fraction  $\Delta_{\mathrm{ph}}/\Delta<20\%$, PPC when  $\Delta_{\mathrm{ph}}/\Delta>80\%$, and IPC region for values between these thresholds~\cite{kamide2010,KO11}.

For the weak Coulomb interaction, Fig.~\ref{f1}(c) shows us that the system predominantly resides in the PPC phase. The photonic fraction $\Delta_{\mathrm{ph}}/\Delta$ consistently exceeds 80$\%$, while the excitonic fraction $\Delta_{\mathrm{eh}}/\Delta$ remains negligible with increasing excitation density. This behavior confirms that when the Coulomb attraction is insufficient to tightly bind electron-hole pairs, the condensate in the SC side is driven almost entirely by the cavity photon field. The effect of mass imbalance is relatively subtle here, however, increasing the hole hopping integral $t^h$ further enhances the photonic character, pushing the $\Delta_{\mathrm{ph}}/\Delta$ ratio even closer to unity. In contrast, the strong coupling regime reveals a rich crossover behavior driven by the interplay between the mass imbalance and excitation density [see Fig.~\ref{f1}(d)]. At low excitation densities, the condensate is purely excitonic in nature $\Delta_{\mathrm{eh}}/\Delta\approx 1$, signifying the formation of a stable EPC. This stability is rooted in the strong local Coulomb attraction, which binds the electron-hole pairs into tight, composite bosons well below the photon band edge. As the density increases, a clear crossover occurs, the excitonic fraction monotonically decreases while the photonic fraction rises, eventually driving the system through the IPC regime and into the PPC phase at high densities. This transition is a manifestation of the saturation of the excitonic component due to Pauli blocking, as discussed in the previous panels, contrasted with the unbounded Bose-stimulated accumulation of photons.

The significant role of mass imbalance in governing the stability of the excitonic phase is highlighted in Fig.~\ref{f1}(d). We observe that the critical density at which the crossover from EPC to IPC occurs, where $\Delta_{\mathrm{ph}}/\Delta$ crosses 20$\%$, significantly depends on $t^h$. For small $t^h$ or strong mass imbalance, the excitonic-like phase occurs up to much higher excitation densities. Indeed, a smaller $t^h$ corresponds to a flatter valence band and a significantly higher DOS. Consequently, as the particle number increases, the chemical potential rises more slowly than the case of light holes. This `pinning' of the chemical potential allows the system to accommodate more particles within the excitonic manifold before energetically accessing the photon-dominated region of the dispersion. Conversely, for large $t^h$, the holes become lighter, and the lower DOS forces a rapid rise in the chemical potential, triggering an earlier onset of the photonic dominance. Thus, strong mass imbalance acts as a stabilizing mechanism for the PCs, extending the EPC phase boundaries against the effects of phase-space filling.

\begin{figure}[h]
\includegraphics[width=0.47\textwidth]{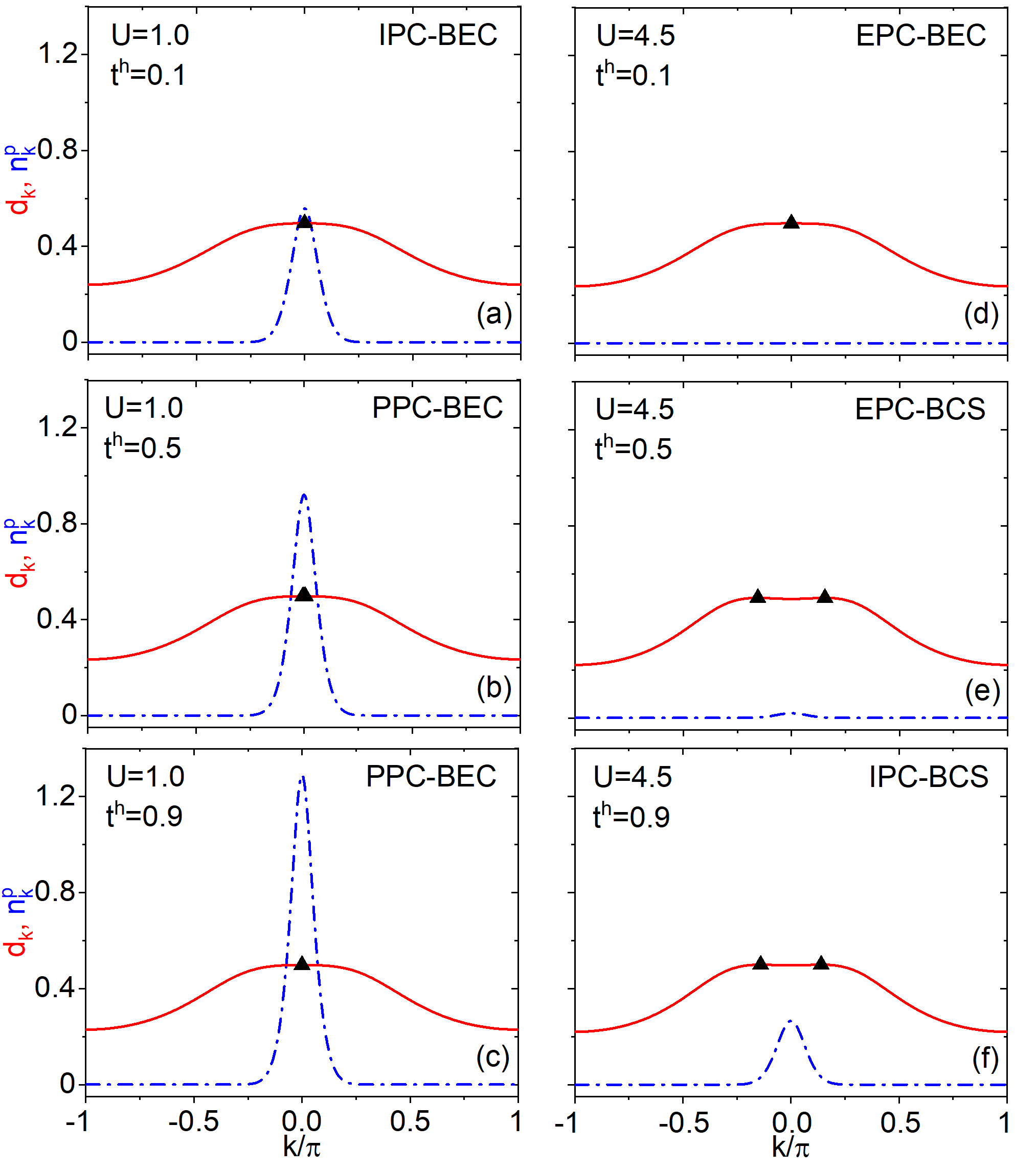}
\caption{Momentum distributions of the electron-hole pair amplitude $d_\mathbf{k}$ (red solid lines) and the photon density $n^p_\mathbf{k}$ (blue dashed lines) in the first Brillouin zone along the $\mathbf{k}=(k,0)$ direction for three different values of $t^{h}$ with $n = 0.1$ at $U=1$ (left column) and $U=4.5$ (right column). The triangle symbols indicate the maximum position of $d_\mathbf{k}$.}
\label{f2}
\end{figure}
 
To distinguish the microscopic nature of the condensed states, specifically whether they exhibit the BCS or BEC type, we analyze the internal structure of the condensate in momentum space. Figure~\ref{f2} depicts the momentum distributions of the electron-hole pair amplitude $d_\mathbf{k}$ and the photon density $n^p_\mathbf{k}$ within the first Brillouin zone along the $\mathbf{k}=(k,0)$ direction for two limits of Coulomb interaction (weak $U=1$, left column, and strong $U=4.5$, right column) by varying the hole hopping integral $t^h$. Note here that, in the SC situation, the bottom of the photon band lies slightly below the excitonic resonance, which enhances the excitonic contribution relative to the photonic one and favors the formation of well-defined bound electron-hole pairs. As a result, the BCS-BEC crossover would be sensitively affected not only by the mass imbalance but also the interlayer Coulomb interaction.

For weak Coulomb attraction, Fig.~\ref{f2}(a)--\ref{f2}(c) show a single broad maximum in $d_\mathbf{k}$ for all hole hoppings $t^{h}$. This robust single-peak structure indicates a BEC-like regime dominated by tightly bound excitons or exciton-polaritons~\cite{PRX.11.011018}. The nature of the condensate varies with the mass ratio. For heavy holes $t^{h}=0.1$, the ground state remains predominantly IPC-BEC [panel (a)]. As $t^{h}$ increases to $0.5$ and $0.9$, a sharp peak in $n^p_\mathbf{k}$ emerges at $k=0$, signifying a transition toward a PPC-BEC, although the electron-hole pairs remain tightly bound in the SC situation [panels (b) and (c)]. For strong Coulomb attraction, $U=4.5$, the influence of mass imbalance becomes even more pronounced. Indeed, at $t^{h}=0.1$, the system retains a single central peak in $d_\mathbf{k}$ with negligible photonic weight, consistent with a deeply bound EPC-BEC state [panel (d)]. However, at intermediate hopping $t^{h}=0.5$, two distinct maxima develop in $d_\mathbf{k}$, demonstrating the emergence of a well-defined Fermi momentum and a crossover toward a BCS-like excitonic condensate, despite the SC gap and negative detuning [panel (e)]. In the large mass imbalance situation, the photonic weight is nearly neglectable and the system settles in EPC. For nearly mass-balanced carriers $t^{h}=0.9$, the two-peak structure persists [panel (f)], and the reinforced photonic weight near $k=0$ indicates that the system forms a IPC in the BCS type, where Coulomb-driven extended pairing dominates while light-matter coupling hybridizes the low-energy quasiparticles. Due to the strong excitonic binding, the system exhibits an EPC-BEC scenario at extremely large mass asymmetry and then EPC-BCS behavior at moderate mass asymmetry. Only when the mass ratio becomes nearly balanced does the condensate acquire a significant photonic component and transition toward a IPC-BCS regime. The combined influence of detuning, mass imbalance, and Coulomb attraction therefore generates a rich sequence of EPC, PPC, and IPC of the PCs with the BCS-BEC crossover.

\begin{figure}[h]
\includegraphics[width=0.45\textwidth]{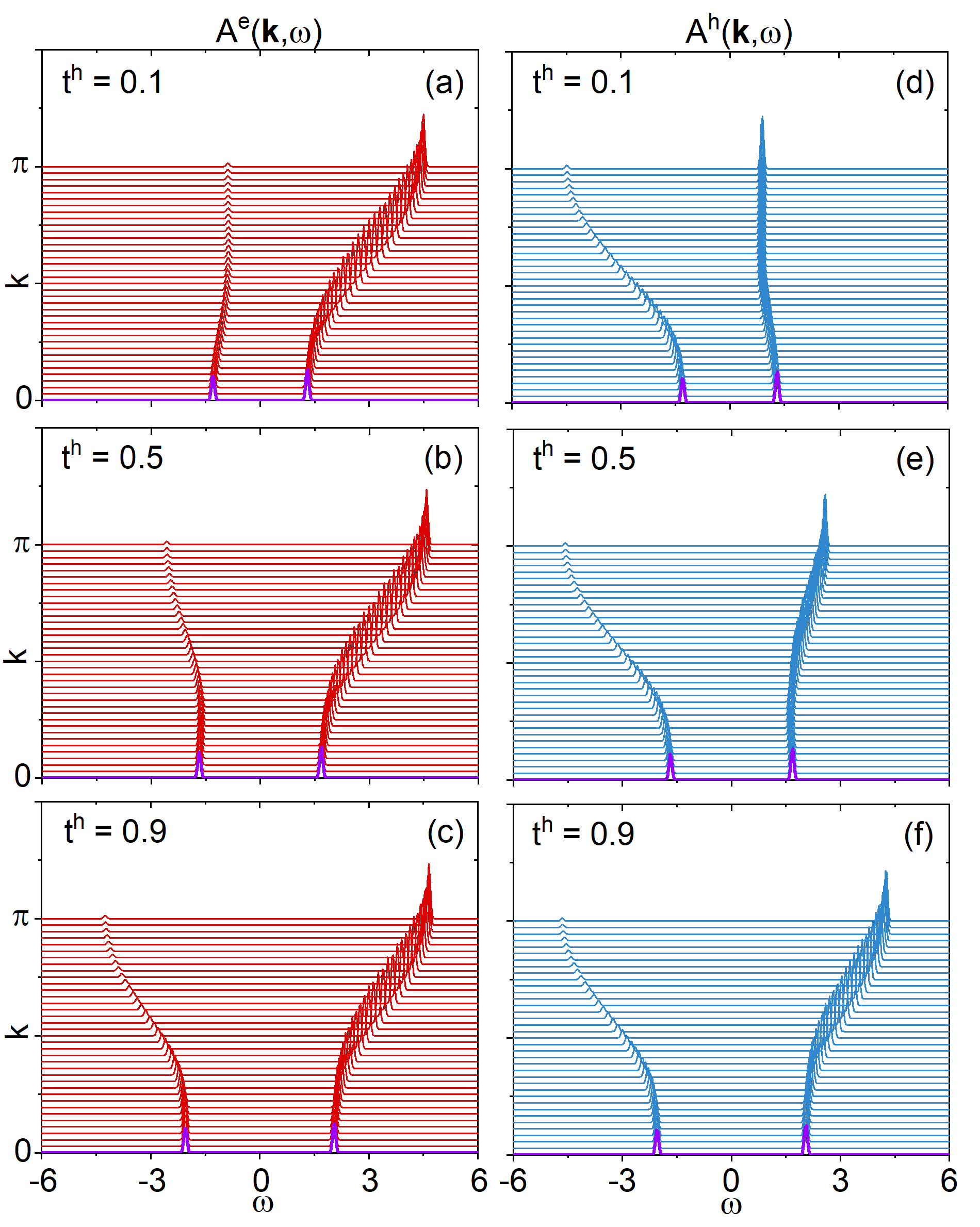}
\caption{{Wave vector--resolved single-particle spectral functions of} electrons $A^{e}(\mathbf{k}, \omega)$ (red lines) and holes $A^{h}(\mathbf{k}, \omega)$ (blue lines) along the $\mathbf{k}=(k,0)$ direction ($k>0$) for the parameter set corresponding to the left column of Fig.~\ref{f2}. The spectra highlighted in purple mark the contributions at the Fermi momentum.}
\label{f3}
\end{figure}

\begin{figure}[t]
\includegraphics[width=0.45\textwidth]{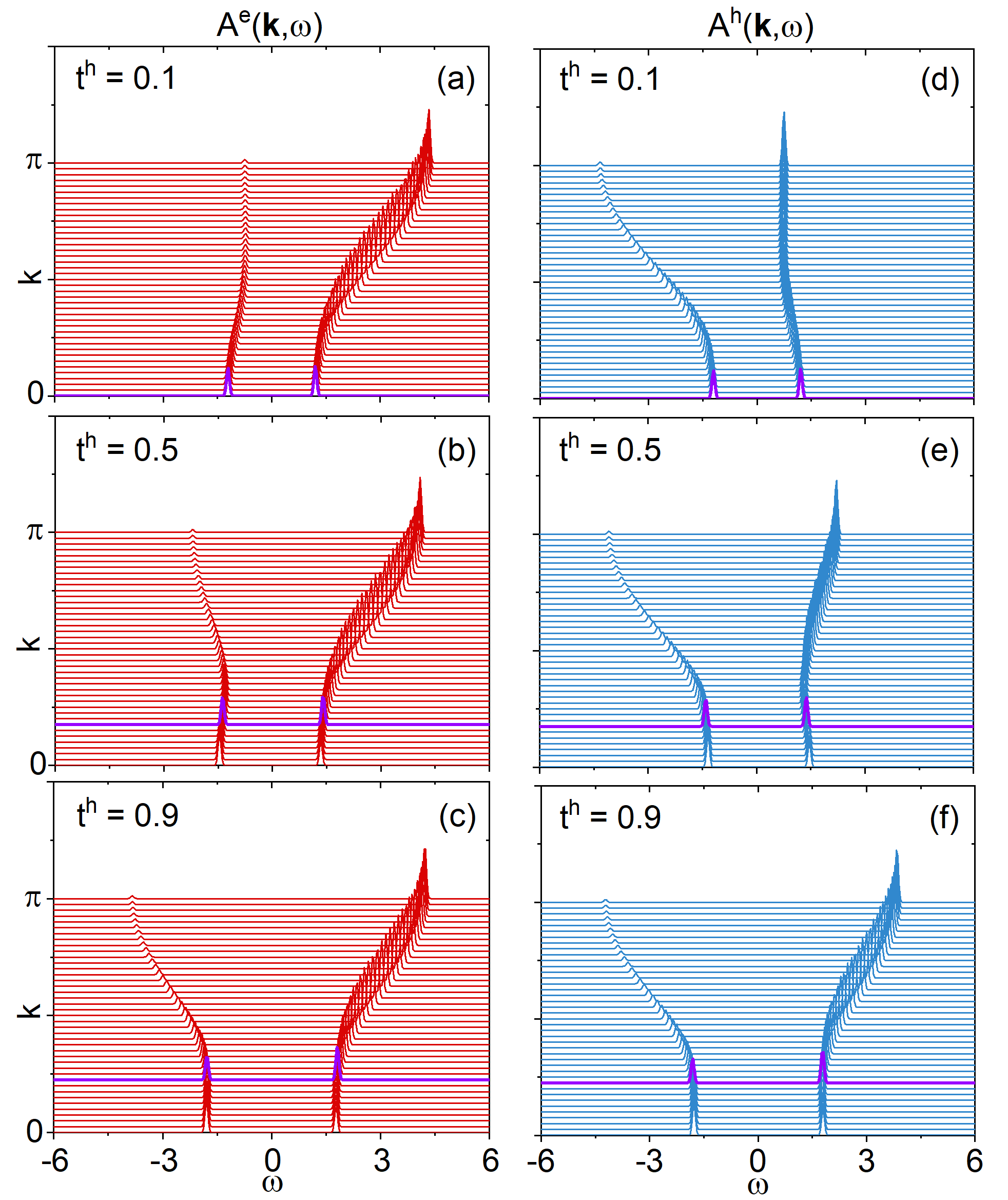}
\caption{{Wave vector--resolved single-particle spectral functions of} electrons $A^{e}(\mathbf{k}, \omega)$ (red lines) and holes $A^{h}(\mathbf{k}, \omega)$ (blue lines) along the $\mathbf{k}=(k,0)$ direction ($k>0$) for the parameter set corresponding to the right column of Fig.~\ref{f2}. The spectra highlighted in purple mark the contributions at the Fermi momentum.}
\label{f4}
\end{figure}

To further elucidate the influence of electron-hole mass imbalance and Coulomb interaction on the PCs, specially the nature of their BCS-BEC crossover, we examine the wave-vector- and frequency-resolved single-particle spectral functions of electrons $A^{e}(\mathbf{k},\omega)$ and holes $A^{h}(\mathbf{k},\omega)$. Figures~\ref{f3} and~\ref{f4} show the spectral intensities along the $\mathbf{k}=(k,0)$ direction ($k>0$) in the first Brillouin zone for the parameter sets with respect to the left and right columns in Fig.~\ref{f2}. Figure~\ref{f3} shows the spectral functions in the weak-coupling regime with $U=1$, where as indicated by the pairing amplitudes in Fig.~\ref{f2}, the system lies deep in the BEC side of the crossover. For strongly imbalanced masses, the valence band becomes nearly flat and the hole spectrum exhibits a narrow, almost vertical branch centered around $k=0$, reflecting the extremely heavy hole mass [Figs.~\ref{f3}(a) and \ref{f3}(d)]. In the normal state, the spectra would consist of a single conduction band for electrons and a single valence band for holes. In contrast, the present spectra exhibit a pronounced redistribution of spectral weight into two hybridized branches for both $A^{e}(\mathbf{k}, \omega)$ and $A^{h}(\mathbf{k}, \omega)$, separated by a sizable hybridization gap. The high accumulation of spectral weight at zero momentum (highlighted in purple) reflects the dominant coherent mixing of electron-hole degrees of freedom at $k=0$ induced by Coulomb attraction and light-matter coupling. The signature specifies the formation of tightly bound electron-hole pairs with suppressed center-of-mass motion, providing direct spectroscopic evidence for the EPC-BEC state identified earlier. At intermediate hopping, the mobility of the hole is enhanced that reinforces redistribution between the electronlike and holelike branches. Nevertheless, the spectra still display the hallmarks of a BEC regime with spectral intensity concentrated near $k=0$ [Figs.~\ref{f3}(b) and \ref{f3}(e)]. Even for nearly balanced masses, the electron and hole dispersions become more symmetric, the quasiparticle features remain molecularlike, with a large hybridization gap and no indication of Fermi momentum and the system remains in the BEC regime because the negative detuning stabilizes tightly bound excitons and suppresses extended Cooper-like pairing [Figs.~\ref{f3}(c) and \ref{f3}(f)]. 

The spectral functions reveal a qualitatively different behavior when the Coulomb attraction is strong. At $U=4.5$, Fig.~\ref{f4} shows that the hole band remains nearly flat and the spectral features continue to reflect a deeply bound, localized excitonic state once $t^{h}=0.1$ [Fig.~\ref{f4}(a$\&$d)]. Despite the large interaction strength, the enormous mass imbalance prevents the formation of a Fermi momentum, and the system remains in the BEC like state. By lowering the mass imbalance, the system undergoes a qualitative change. Two dispersive quasiparticle branches appear in both the electron and hole spectra, bending toward each other near a well-defined momentum [Fig.~\ref{f4}(b) and \ref{f4}(e)]. These features mark the emergence of extended electron-hole pairing, signaling the transition to a BCS regime. This interpretation is supported by the two-peak structure in the pairing amplitude in Fig.~\ref{f2}(e), which indicates the appearance of a Fermi momentum. For nearly mass-balanced carriers, at $t^{h}=0.9$, the quasiparticle dispersions evolve into well-defined Bogoliubov-like upper and lower bands with strong spectral weight on both branches [Fig.~\ref{f4}(c) and \ref{f4}(f)]. The hybridization gap continues to enhance, consistent with the development of particle-hole coherence typical of a BCS condensate. The reduced effective mass imbalance allows both carriers to participate equally in electron-hole pairing, producing the BCS state [c.f. Fig.~\ref{f2}(f)]. 

\begin{figure}[b]
\includegraphics[width=0.49\textwidth]{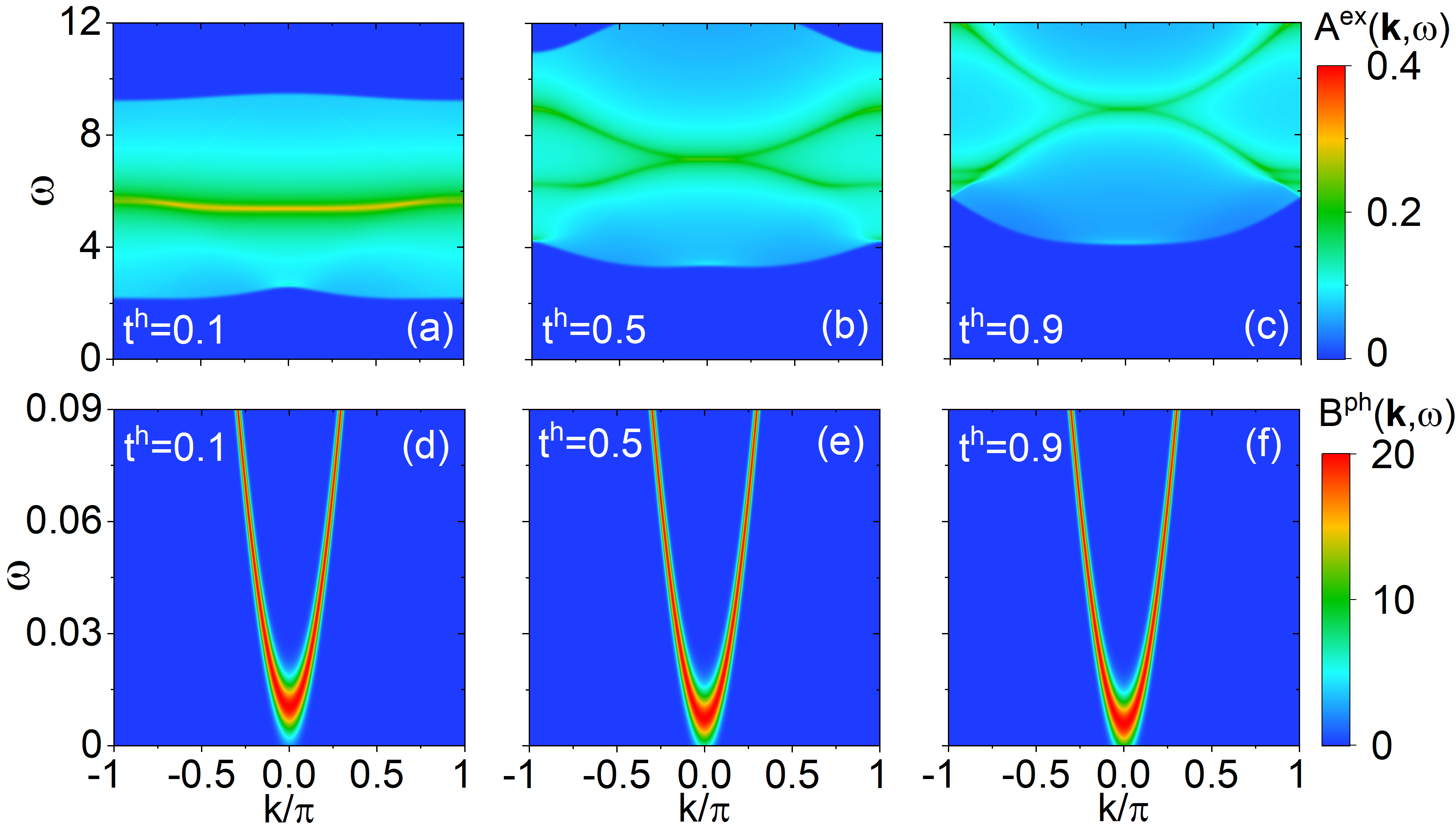}
\caption{Intensity plot of the luminescence functions for excitonic polarization $A^{\textrm{ex}}(\mathbf{k}, \omega)$ (top) and for cavity photon mode $B^{\textrm{ph}}(\mathbf{k}, \omega)$ (bottom) along the $\mathbf{k}=(k,0)$ direction for the parameter set corresponding to the left column of Fig.~\ref{f2}.}
\label{f5}
\end{figure}

To inspect deeper insight into the quasiparticle excitation spectrum in the SC state, we examine the signatures of the luminescence functions for the excitonic polarization $A^{\textrm{ex}}(\mathbf{k}, \omega)$ and the cavity photon mode $B^{\textrm{ph}}(\mathbf{k}, \omega)$ addressed respectively in Eqs.~\eqref{eq16} and~\eqref{eq17} for the parameters settled in Fig.~\ref{f2}. Fig.~\ref{f5} presents the luminescence spectra for weak Coulomb interaction, $U=1.0$. The top row shows that the excitonic-polarization spectrum exhibits pronounced depletion of excitonic spectral weight emerge at low energies, indicating the coherent electron-hole pairing that increases as lowering the mass imbalance. In the situation of small $t^h$, the spectrum is nearly flat, reflecting tightly bound excitons with very weak dispersion [Fig.~\ref{f5}(a)]. As $t^{h}$ increases, the excitonic mode acquires stronger dispersion and hybridizes more visibly with the higher-energy band [Fig.~\ref{f5}(b) and \ref{f5}(c)]. This evolution directly reflects the enhancement of hole mobility, which reduces the mass imbalance and promotes the formation of spatially extended electron-hole pairs. The smearing of the excitonic peak signals a crossover from tightly bound excitons to more delocalized, overlapping polaritonic excitations. 

Across all mass ratios, the photonic luminescence spectra $B^{\mathrm{ph}}(\mathbf{k},\omega)$ show a weak dependence on $t^{h}$, remaining a sharp, parabolic mode like the bare cavity dispersion [Fig.~\ref{f5}(d) and \ref{f5}(f)]. Nevertheless, a small but systematic shift downward to zero frequency of the spectra with increasing $t^{h}$ is visible, consistent with the fact that light-matter hybridization becomes more effective once the excitonic dispersion broadens and the electronic subsystem enters a more delocalized regime. The downward shift of the photon spectral intensity with increasing $t^h$ further highlights the enhanced photonic contribution to the matter-light system, thereby stabilizing the photonic character of the PC state.

\begin{figure}[h]
\includegraphics[width=0.49\textwidth]{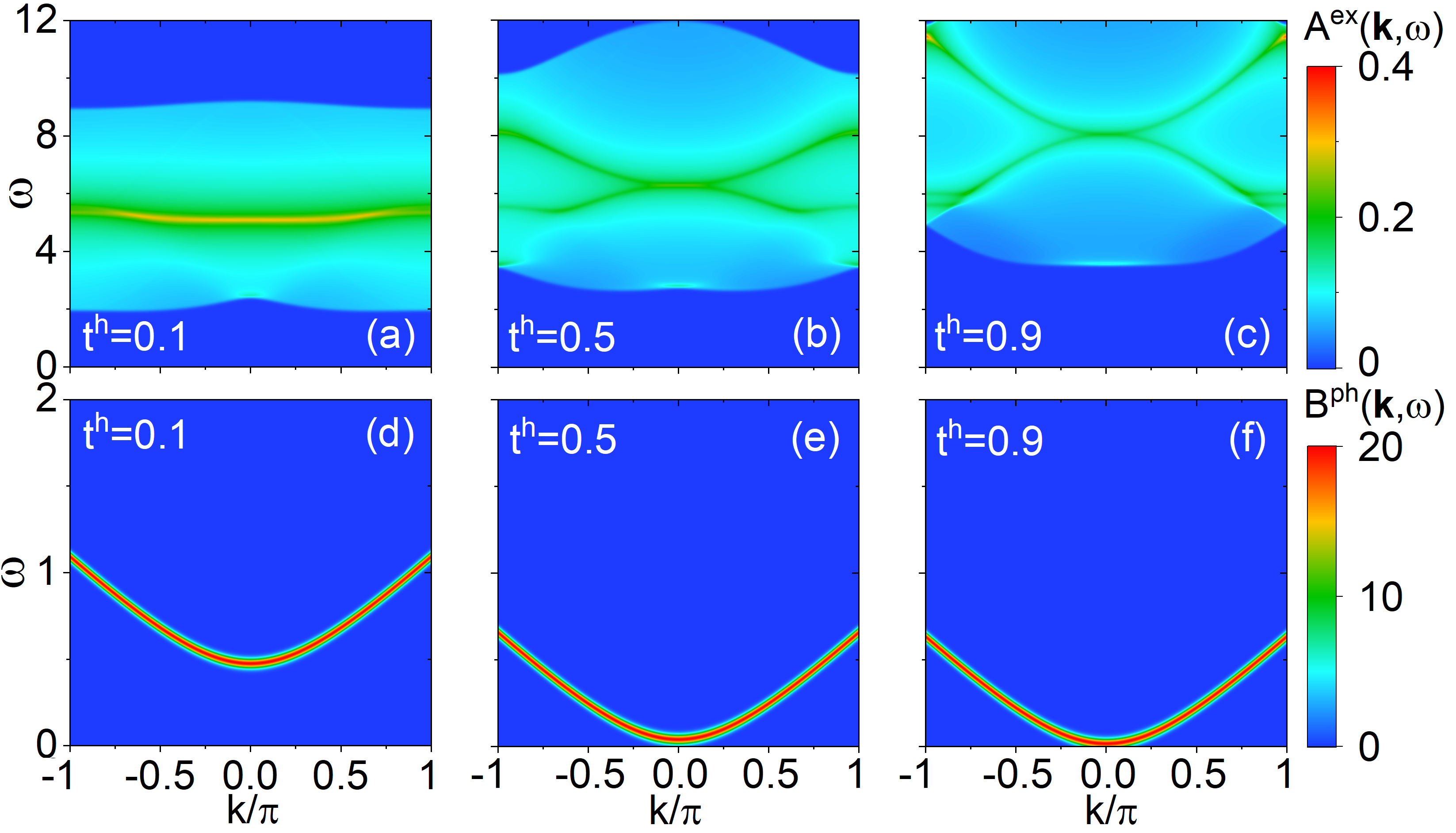}
\caption{Intensity plot of the luminescence functions for excitonic polarization $A^{\textrm{ex}}(\mathbf{k}, \omega)$ (top) and for cavity photon mode $B^{\textrm{ph}}(\mathbf{k}, \omega)$ (bottom) along the $\mathbf{k}=(k,0)$ direction for the parameter set corresponding to the right column of Fig.~\ref{f2}.}
\label{f6}
\end{figure}

In the situation of large Coulomb interaction, $U=4.5$, signatures of the excitonic-polarization luminescence $A^{\mathrm{ex}}(\mathbf{k},\omega)$ and the photon luminescence $B^{\mathrm{ph}}(\mathbf{k},\omega)$ are illustrated in Fig.~\ref{f6}. Similar to the case of small Coulomb interaction,  $A^{\mathrm{ex}}(\mathbf{k},\omega)$ displays a sharp, intense, and weakly dispersive resonance at small $t^h$ [Fig.~\ref{f6}(a)]. As $t^{h}$ increases, the excitonic resonance becomes broader and more dispersive [Figs.~\ref{f6}(b) and \ref{f6}(c)], signifing a crossover from tightly bound excitons to more extended, overlapping electron-hole pairs. The photon luminescence $B^{\mathrm{ph}}(\mathbf{k},\omega)$ behaves in a different manner. Indeed, at small $t^{h}$ the photon mode lies at relatively large $\omega$, reflecting the energy mismatch between the heavy-mass excitonic sector and the cavity mode [Fig.~\ref{f6}(d)]. Increasing $t^{h}$ shifts the photon dispersion downward toward $\omega\simeq 0$ [Figs.~\ref{f6}(e) and \ref{f6}(f)], indicating that the effective detuning between matter and light decreases as the carrier masses become more balanced. This shift enhances the overlap between the photon mode and the exciton-polariton resonance, thereby increasing the photonic contribution to the condensate. The approach of the photon branch toward lower energies is therefore a direct signature of the growing role of the cavity field in the PCs as mass imbalance is reduced.

\subsection{Semimetal side}

\begin{figure}[h]
\includegraphics[width=0.47\textwidth]{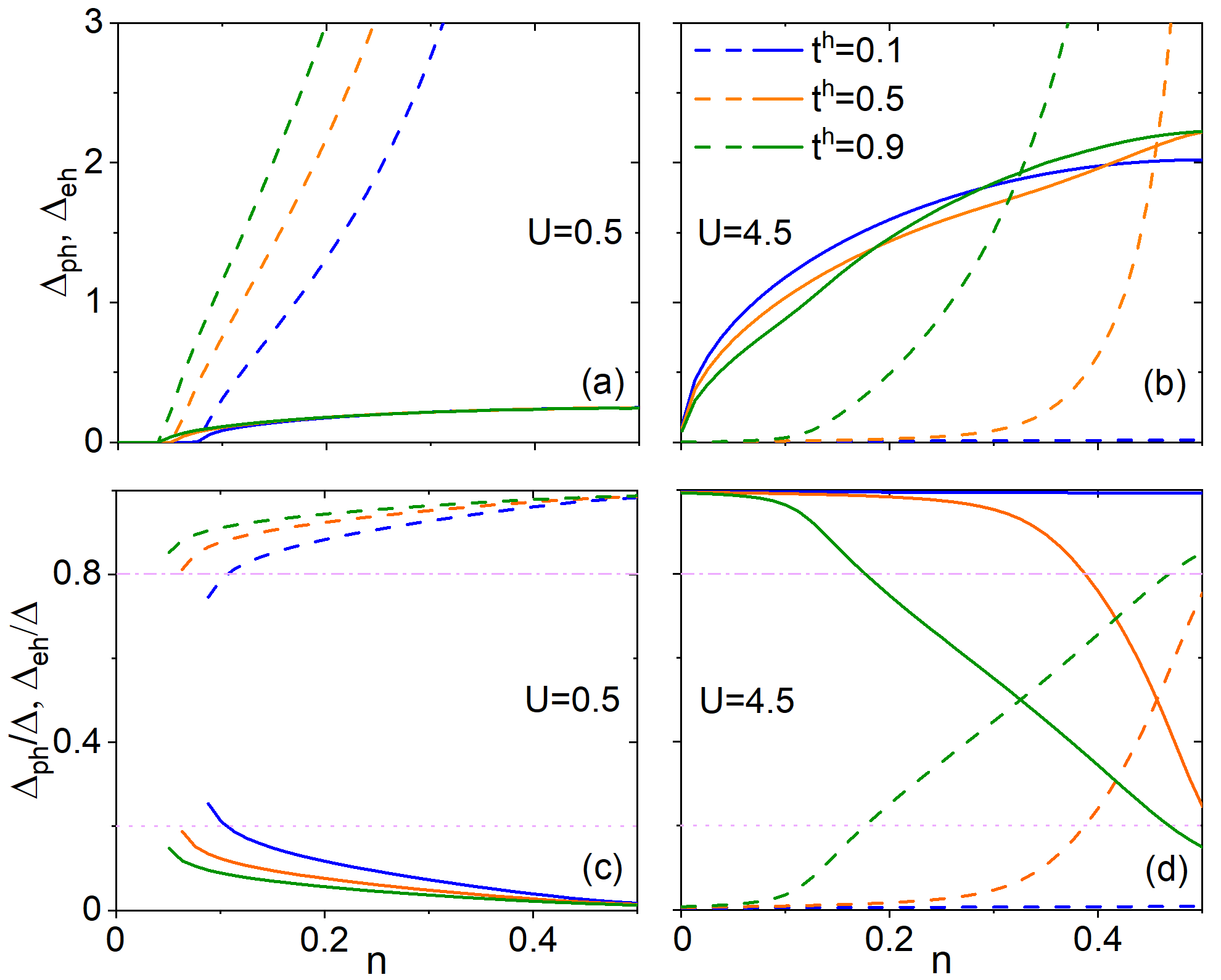}
\caption{Condensate order parameters as functions of the excitation density $n$ for different hole hopping amplitudes $t^h$ at detuning parameter $d=1$. The top row shows the excitonic $\Delta_{\mathrm{eh}}$ (solid lines) and the photonic $\Delta_{\mathrm{ph}}$ (dashed lines), while the bottom row displays their relative weights in the condensates for Coulomb interaction (a) and (c) $U=0.5$ and (b) and (d) $U=4.5$. The horizontal dotted and dashed-dotted lines mark the fractions of 20$\%$ and 80$\%$, respectively.}
\label{f7}
\end{figure}

Next, we examine signatures of the PCs in the SM side by considering the positive detuning case with $d=1$. In this situation, the bare electron and hole bands overlap, and the cavity photon mode lies energetically above the electronic continuum. Figure~\ref{f7} summarizes the dependence of the excitonic $\Delta_{\mathrm{eh}}$ and photonic $\Delta_{\mathrm{ph}}$ condensate order parameters and their ratios $\Delta_{\mathrm{eh}}/\Delta$ and $\Delta_{\mathrm{ph}}/\Delta$ on the excitation density $n$.

Due to the positive detuning, the cavity photon mode lies energetically above the electron–hole continuum, such that at low excitation densities the chemical potential resides entirely within the matter sector. For weak Coulomb interaction, coherent condensation is therefore possible only once the chemical potential approaches the photon energy. In this regime, light-matter coupling dominates the instability and favors macroscopic occupation of the photon mode. As shown in Fig.~\ref{f7}(a), both condensate order parameters exhibit a clear threshold behavior, remaining vanishingly small up to a critical excitation density. This reflects the fact that, despite the SM band overlap, the combined DOS and interaction strength are insufficient to sustain spontaneous coherence at very low densities. Once the threshold is exceeded, the photonic order parameter $\Delta_{\mathrm{ph}}$ increases rapidly and quickly surpasses the excitonic component $\Delta_{\mathrm{eh}}$. Consistently, Fig.~\ref{f7}(c) demonstrates that the photonic fraction $\Delta_{\mathrm{ph}}/\Delta$ exceeds $80\%$ already at low densities for all values of the hole hopping $t^{h}$, identifying a predominantly photonic polariton condensate. The weak dependence on $t^{h}$ highlights the minor role of Coulomb-induced electron–hole binding in this regime, where condensation is primarily driven by light-matter hybridization.

In contrast, the strong-coupling regime exhibits qualitatively different behavior [Fig.~\ref{f7}(b)]. Here, the enhanced Coulomb attraction induces electron-hole pairing already at extremely low excitation densities, driving the system into an excitonic-polariton condensate. In this limit, the chemical potential remains well below the photon energy, and the condensate is dominated by $\Delta_{\mathrm{eh}}$, reflecting an excitonic instability rooted in the SM background~\cite{Mo61,Kno63,JRK67,PRL.19.439,HR68b}. Upon increasing the excitation density, the chemical potential gradually approaches the photon level, leading to a growth of the photonic order parameter and a crossover toward a photonic-dominated condensate. As illustrated in Fig.~\ref{f7}(d), this redistribution of condensate weight is strongly controlled by the hole hopping amplitude. Smaller $t^{h}$ stabilizes the excitonic fraction over a broad density window, whereas larger $t^{h}$ accelerates the transfer of condensate weight to the photonic sector. Physically, a heavier hole mass enhances the electronic DOS, causing the chemical potential to increase more slowly with density and thereby delaying resonance with the photon mode. As a result, the system remains in the EPC or IPC regime up to relatively high excitation densities. Conversely, for lighter holes, the reduced DOS leads to a more rapid increase of the chemical potential, bringing the system into resonance with the photon mode at lower densities and promoting an earlier crossover to a IPC or PPC regime.

\begin{figure}[t]
\includegraphics[width=0.45\textwidth]{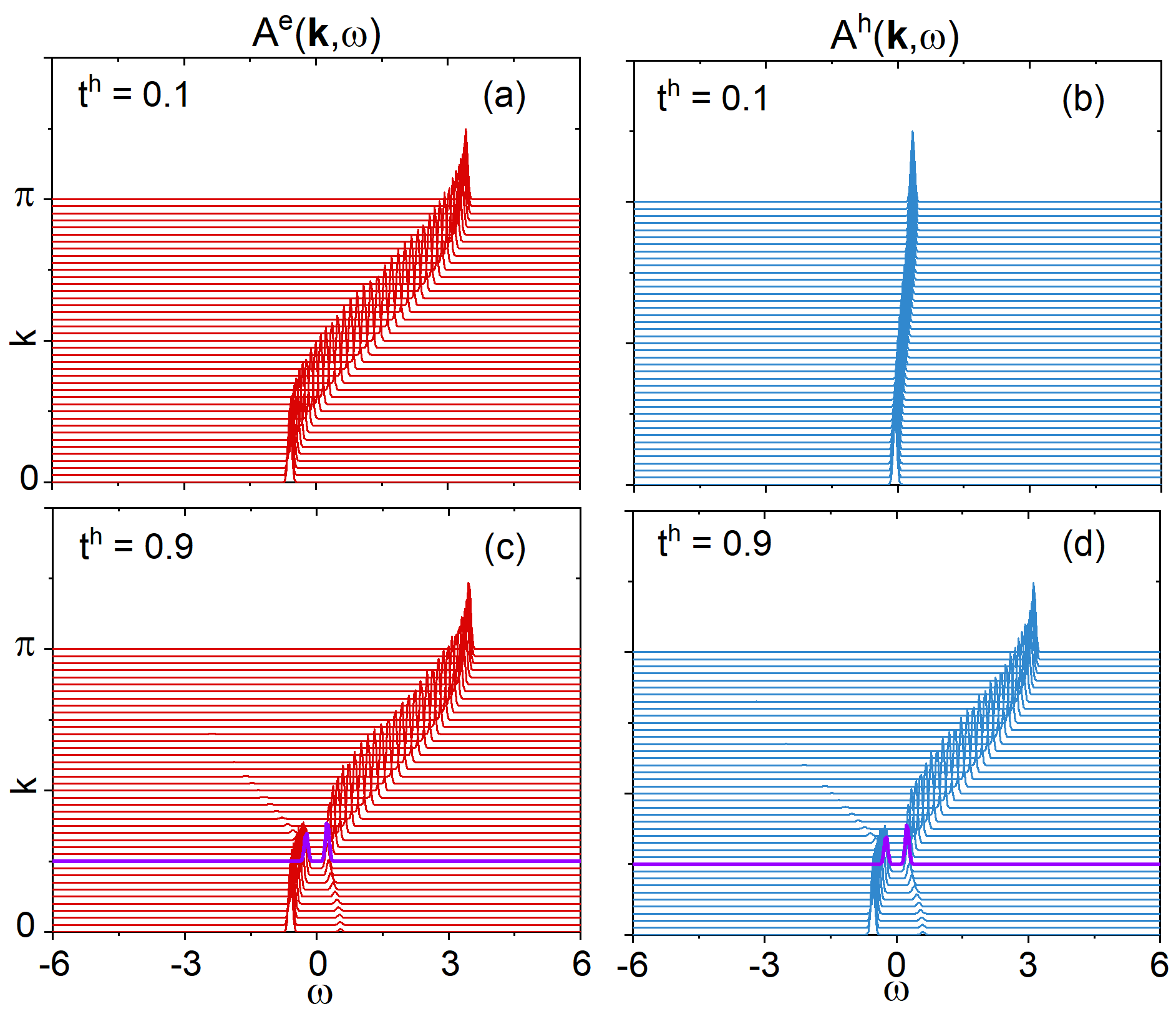}
\caption{{Wave-vector-resolved single-particle spectral functions of} electrons $A^{e}(\mathbf{k}, \omega)$ (red lines) and holes $A^{h}(\mathbf{k}, \omega)$ (blue lines) along the $\mathbf{k}=(k,0)$ direction ($k>0$) for $d = 1$, $n = 0.05$ and $U=0.5$. The spectra highlighted in purple mark the contributions at the Fermi momentum.}
\label{f8}
\end{figure}

To investigate the influence of mass imbalance and Coulomb interaction on the PCs and their BCS-BEC crossover in the SM regime, we analyze the single-particle spectral functions for electrons, $A^{e}(\mathbf{k},\omega)$, and holes, $A^{h}(\mathbf{k},\omega)$ along the $\mathbf{k}=(k,0)$ direction ($k>0$) at $d=1$ and $n=0.05$ for some values of the hole hopping integral $t^h$ and two typical limits of the Coulomb interaction. For the weak Coulomb interaction, $U=0.5$, Fig.~\ref{f8} reveals that the single-particle spectral functions exhibit no energy gap at small hole hopping integral $t^h=0.1$ [panels~(a) and (b)]. This behavior stems from the combination of an extremely large mass imbalance, where holes are nearly localized, and a weak attractive potential that is insufficient to bind electron-hole pairs. Consequently, the electron and hole bands remain overlapping without gap formation. Additionally, due to the large positive detuning inherent to the SM regime, the photon band lies far above the Fermi level, suppressing electron-hole-photon hybridization. Thus, the system remains in a normal, noncondensed state. However, reducing the mass imbalance by increasing $t^{h}$ enhances hole mobility, which promotes exciton formation. As seen in the lower panels (c) and (d) of Fig.~\ref{f8} for $t^h=0.9$, an energy gap develops around the Fermi level in both electron and hole spectral functions. The spectral weight near the Fermi level is distributed almost uniformly at finite momenta, a signature characteristic of a BCS-like condensate where pairing occurs at the Fermi surface.

\begin{figure}[t]
\includegraphics[width=0.45\textwidth]{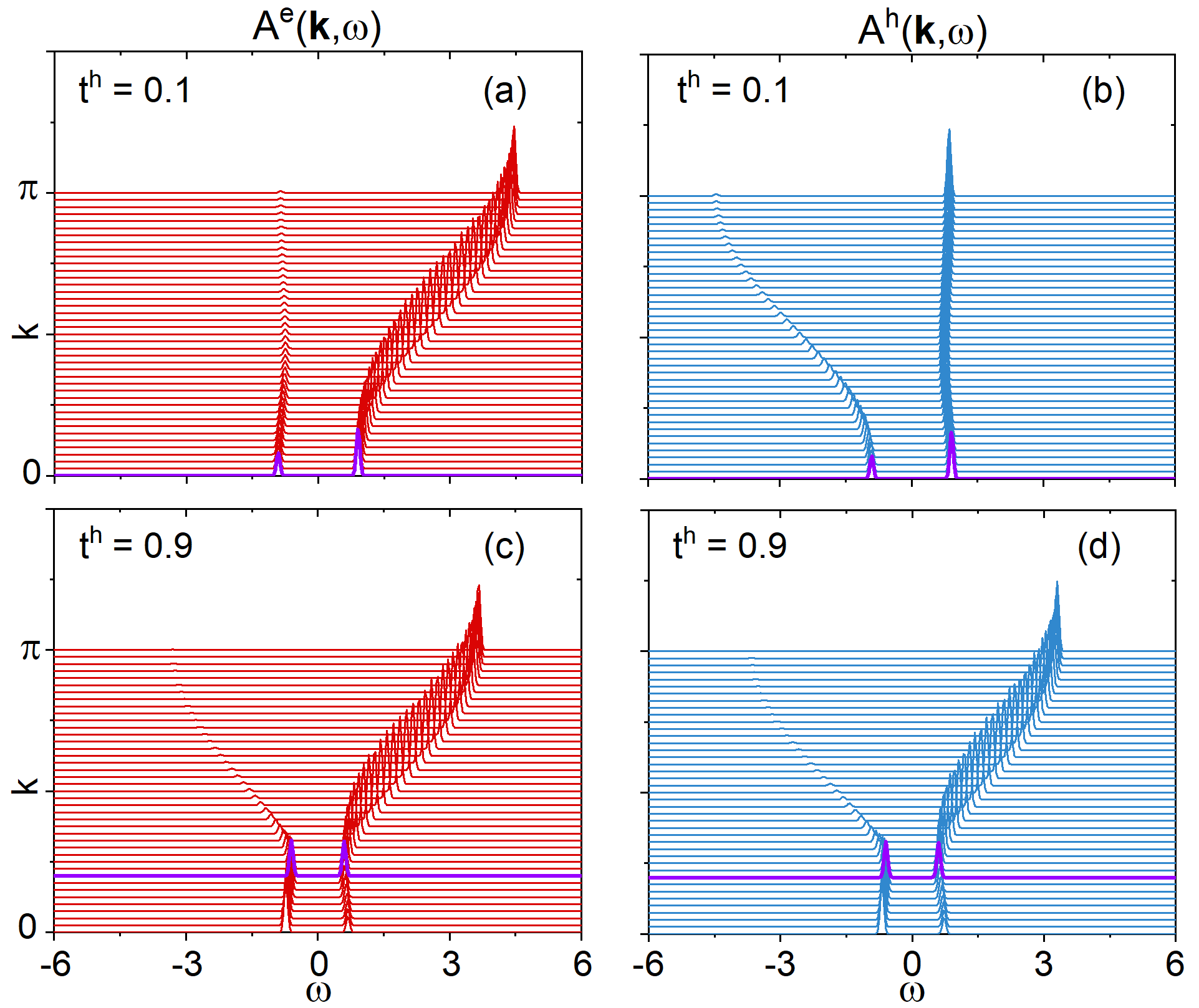}
\caption{{Wave-vector-resolved single-particle spectral functions of} electrons $A^{e}(\mathbf{k}, \omega)$ (red lines) and holes $A^{h}(\mathbf{k}, \omega)$ (blue lines) along the $\mathbf{k}=(k,0)$ direction ($k>0$) for $d = 1$, $n = 0.05$ and $U=4.5$. The spectra highlighted in purple mark the contributions at the Fermi momentum.}
\label{f9}
\end{figure}

In the case of strong Coulomb interaction, at $U=4.5$, the spectral functions are shown in Fig.~\ref{f9}. At large mass imbalance, e.g. $t^h=0.1$, a distinct energy gap appears around the Fermi level [see Fig.~\ref{f9}(a) and (b)]. Crucially, the single-particle spectral weight for both electrons and holes becomes strongly concentrated near momentum $k=0$. This concentration indicates strong electron-hole pairing and the formation of tightly bound excitons, characteristic of a BEC-type state where the center-of-mass motion of the pairs is small. As a result, the system stabilizes in BEC-type condensates. Reducing the mass imbalance in this strong coupling regime alters the spectral distribution. The spectral weight in the valence and conduction bands becomes comparable at finite momenta [see Fig.~\ref{f9}(c) and (d)]. This redistribution indicates that the condensate evolves into the BCS-type state, where electron-hole pairs are formed at the Fermi surface. These results demonstrate that, for a given strong Coulomb interaction, tuning the mass imbalance drives a BCS-BEC crossover within the excitonic condensate or the PCs. Unlike the SC case discussed previously, where the gap is intrinsic to the band structure, here the gap is purely interaction-driven. In the strong coupling limit, the induced gap in the SM becomes so large that the system effectively mimics a wide-gap semiconductor, allowing for a smooth crossover between BCS and BEC limits controlled by the kinetic energy scales of the carriers.

Comparing the SM and SC regimes highlights the fundamentally different ways in which they undergo the BCS-BEC crossover in the impacts of mass imbalance and Coulomb interaction. In the weak interaction limit, the SC system remains firmly on the BEC side for all values of the mass imbalance, reflecting the stability of tightly bound excitons across the entire range of $t^h$. By contrast, in the SM regime the system evolves from a normal state to a BCS-type condensate as the hole mobility increases, demonstrating that itinerancy of both carriers is essential for Cooper-like pairing. At strong Coulomb attraction, however, the two regimes exhibit qualitatively similar behavior with both developing a continuous BEC-BCS crossover as the mass imbalance is reduced, underscoring the dominant role of carrier mobility in enabling extended electron-hole pairing.

\begin{figure}[h]
\includegraphics[width=0.48\textwidth]{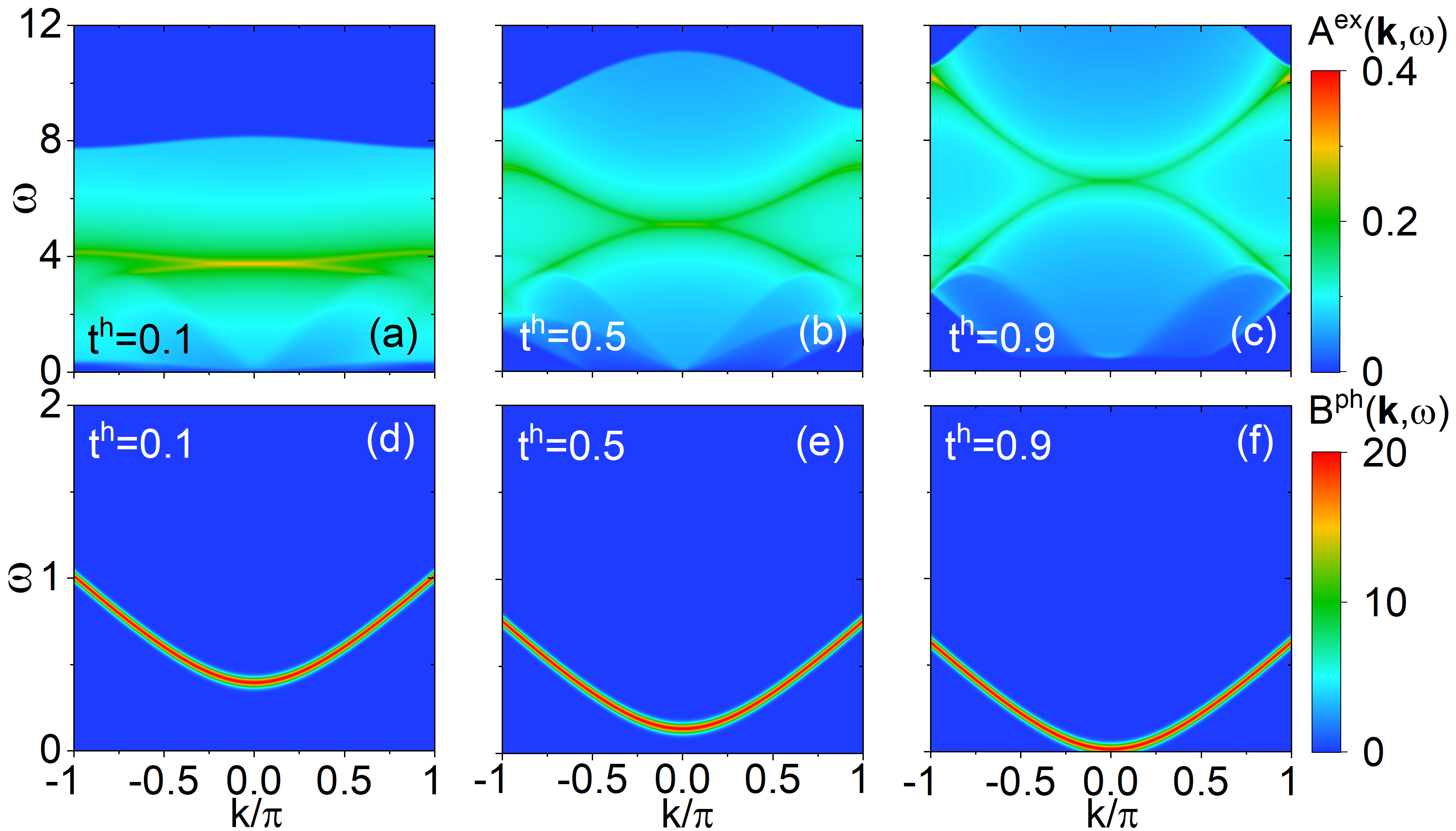}
\caption{Intensity plot of the luminescence functions for excitonic polarization $A^{\textrm{ex}}(\mathbf{k}, \omega)$ (top) and for cavity photon mode $B^{\textrm{ph}}(\mathbf{k}, \omega)$ (bottom) along the $\mathbf{k}=(k,0)$ direction at $d=1.0$, $n=0.05$, and $U=0.5$.}
\label{f10}
\end{figure}

\begin{figure}[h]
\includegraphics[width=0.48\textwidth]{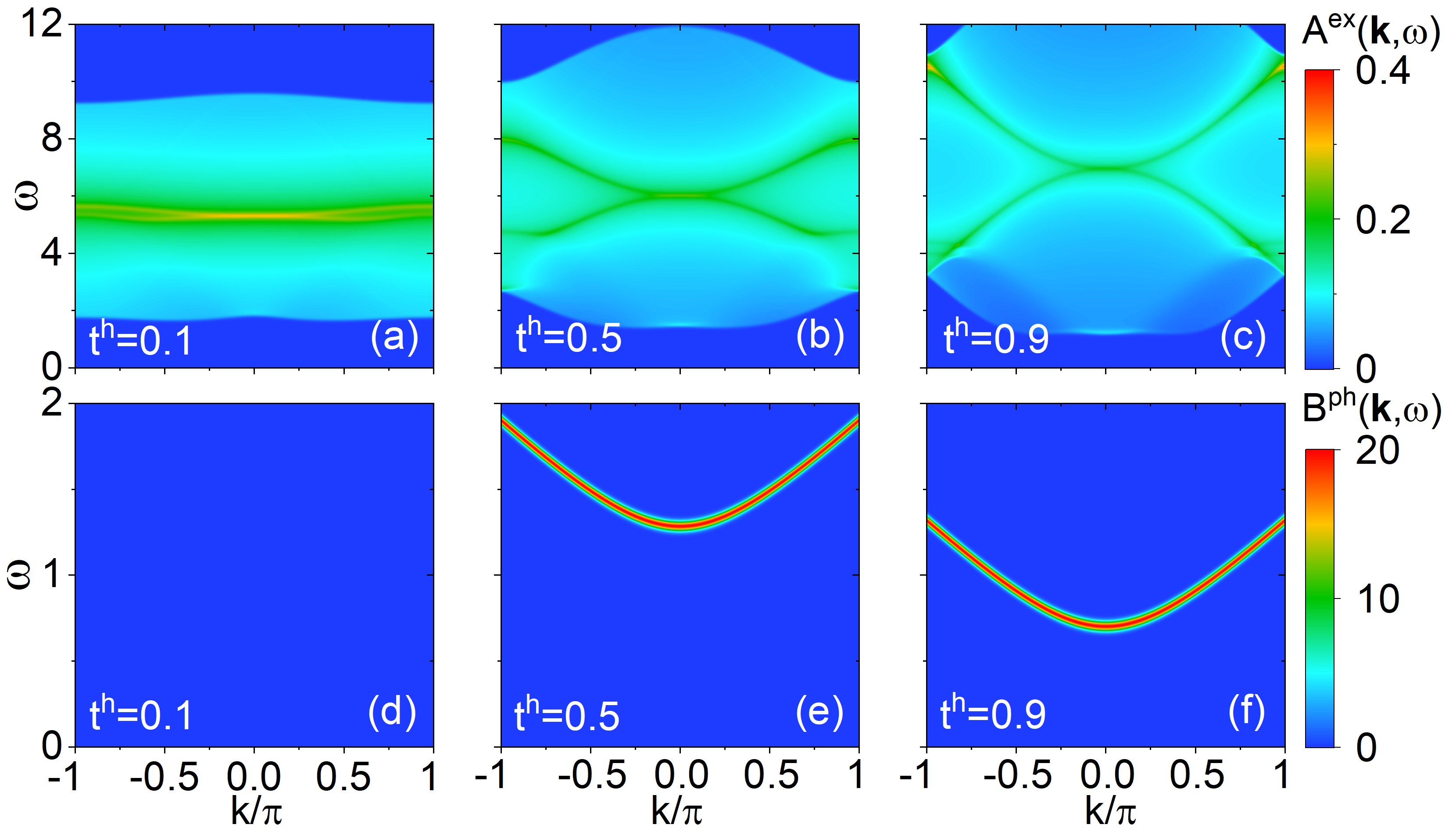}
\caption{Intensity plot of the luminescence functions for excitonic polarization $A^{\textrm{ex}}(\mathbf{k}, \omega)$ (top) and for cavity photon mode $B^{\textrm{ph}}(\mathbf{k}, \omega)$ (bottom) along the $\mathbf{k}=(k,0)$ direction at  $d=1.0$, $n=0.05$, and $U=4.5$.}
\label{f11}
\end{figure}

Figures~\ref{f10} and \ref{f11} present the momentum- and frequency-resolved luminescence spectra of the excitonic polarization $A^{\mathrm{ex}}(\mathbf{k},\omega)$ and the cavity photon mode $B^{\mathrm{ph}}(\mathbf{k},\omega)$ at low excitation density $n=0.05$, for weak  $U=0.5$ and strong $U=4.5$ Coulomb interactions, respectively. For weak Coulomb interaction, the system remains in a normal, noncondensed state at low and intermediate hole hopping amplitudes $t^{h}=0.1$ and $0.5$ [see Fig.~\ref{f7}(a)]. This behavior is directly reflected in the excitonic polarization, where $A^{\mathrm{ex}}(\mathbf{k},\omega)$ exhibits finite spectral weight continuously extending down to low energies around $\mathbf{k}\simeq 0$, indicating no stable excitonic bound state is formed [Fig.~\ref{f10}(a) and (b)]. Only when $t^{h}$ becomes sufficiently large $t^{h}=0.9$ does a pronounced depletion of excitonic spectral weight emerge at low energies, signaling the onset of coherent electron-hole pairing [Fig.~\ref{f10}(c)]. In this regime, the photon luminescence simultaneously shifts toward the chemical potential [see Fig.~\ref{f10}(f) and compare with Fig.~\ref{f10}(d) and \ref{f10}(e)], indicating that condensation is predominantly driven by the photonic component. The resulting ordered phase is therefore identified as PPC.

The situation changes qualitatively at strong Coulomb interaction illustrated in Fig.~\ref{f11}. At small $t^{h}$, the excitonic polarization develops a sharp, nearly dispersionless low-energy feature centered around $\mathbf{k}=0$ [Fig.~\ref{f11}(a)]. This strong concentration of spectral weight reflects the formation of tightly bound electron-hole pairs with a small spatial extent, characteristic of a BEC type of excitonic condensate. The photon luminescence remains well separated from the chemical potential in this regime [Fig.~\ref{f11}(d)], confirming that the condensate is excitonic in nature rather than photonic. As $t^{h}$ increases, the excitonic spectral weight becomes progressively more dispersive and spreads over a wider momentum range [Fig.~\ref{f11}(b) and (c)]. This redistribution indicates enhanced carrier itinerancy and the emergence of momentum-selective pairing, signaling a crossover from tightly bound excitons to extended Cooper-like electron-hole pairs. Despite this evolution, the photon spectrum stays far from resonance [Fig.~\ref{f11}(e) and (f)], demonstrating that the crossover occurs entirely within the excitonic sector. The resulting phase is thus identified as a BCS-type EPC.

\subsection{Ground-state phase diagram}

\begin{figure}[h]
\includegraphics[width=0.47\textwidth]{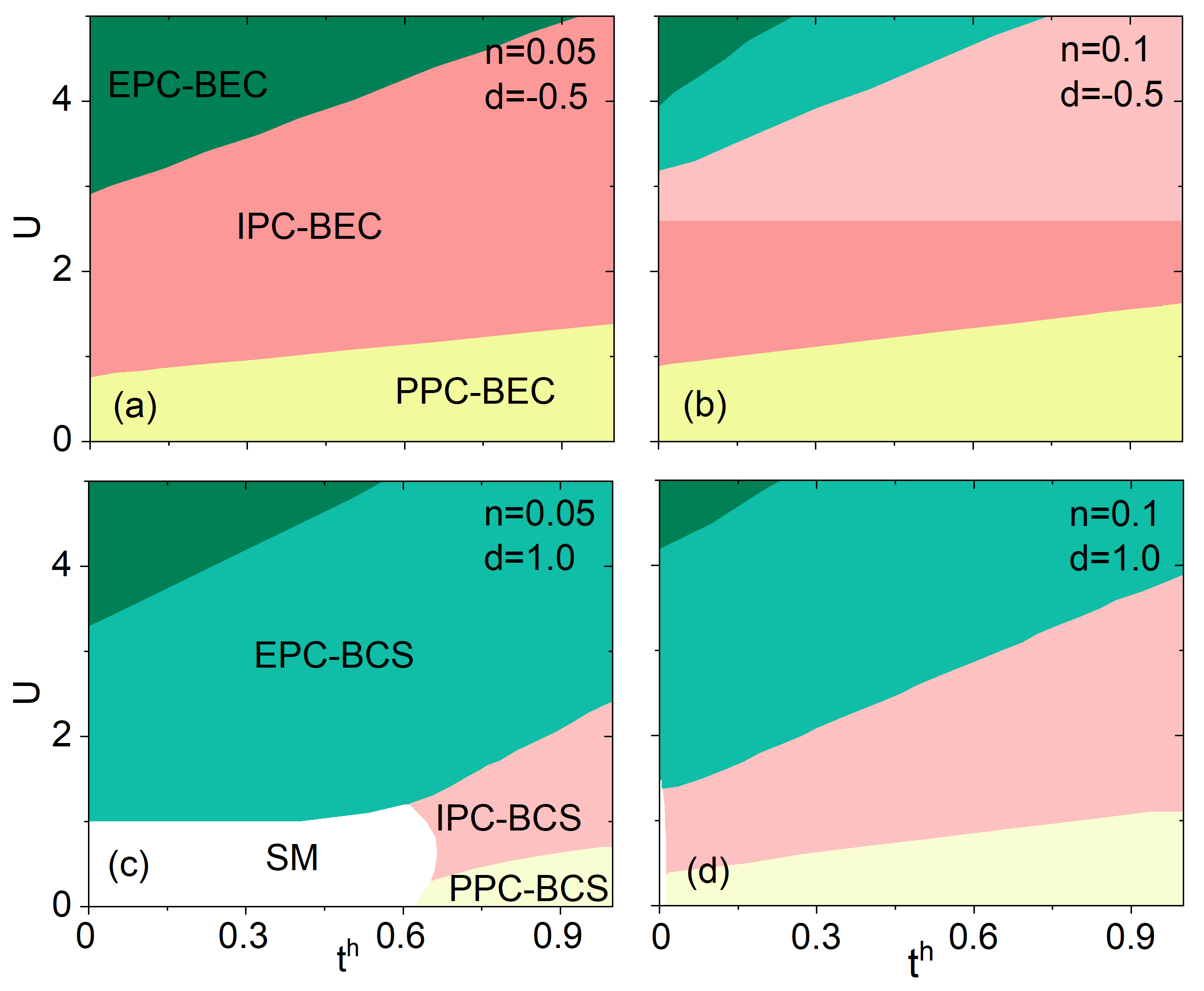}
\caption{Ground-state phase diagram of the mass-imbalance electron-hole-photon system in equilibrium microcavity in the $U$-$t^{h}$ plane in SC, $d=-0.5$, and SM, $d=1$, sides for two different values of excitation density $n$. {The EPC stabilizing in the BCS/BEC type is indicated in light/dark green. The IPC in the BCS/BEC type is shown in light/dark pink while the PPC in BCS/BEC type is marked in light/dark yellow. The white area denotes the normal SM state.}}
\label{f12}
\end{figure}

To address explicitly the competition of the condensation states and their BCS-BEC crossover in microcavity, we summarize the ground-state phase structure of the electron-hole-photon system in the $U$-$t^{h}$ plane in Fig.~\ref{f12} for both the SC (at $d=-0.5$, top row) and SM (at $d=1$, bottom row) regimes at total excitation densities $n=0.05$ (left column) and $n=0.1$ (right column). In the SC regime, Fig.~\ref{f12}(a) and (b) show that the system stabilizes a tightly bound EPC, IPC, or PPC state over a broad range of interaction strengths and mass ratios. At low excitation density $n=0.05$, the condensate remains on the BEC type across nearly the entire parameter space, reflecting the robustness of localized excitons for a positive band gap [panel (a)]. As the excitation density is raised to $n=0.1$, the phase diagram exhibits all condensate types including EPC, IPC, and PPC with a clear BEC-BCS crossover [panel (b)]. At large mass imbalance or small $t^{h}$, the hole mobility is strongly suppressed and the condensates are overwhelmingly BEC-like. In this regime, the strong Coulomb attraction stabilizes an EPC-BEC state, while at intermediate $U$ the matter-light hybridization produces an IPC-BEC state. For weak interaction, the excitonic binding collapses and the condensate becomes photoniclike, yielding the PPC-BEC region. These boundaries reflect the balance between Coulomb-driven local binding and photon-mediated hybridization. Reducing the mass imbalance enhances hole itinerancy that promotes the formation of spatially extended Cooper-like electron-hole pairs. As a result, both EPC and IPC phases undergo a continuous BEC-BCS crossover, and the EPC-BEC and IPC-BEC regions at low $t^{h}$ evolve into EPC-BCS and IPC-BCS at larger $t^{h}$. The critical $t^{h}$ required for this crossover of the EPC increases with $U$ since stronger Coulomb attraction favors tightly bound excitons. In sharp contrast, once the system enters the IPC regime, the substantial light-matter hybridization weakens the sensitivity of the electron-hole pairing character to the degree of mass imbalance and the BCS-BEC crossover of the IPC is nearly independent of $t^{h}$. 

The SM regime with the detuning $d=1$ displays qualitatively different behavior [see Fig.~\ref{f12}(c) and (d)]. Here, the negative band gap reduces the binding strength of electron-hole pairs at weak interaction, and the system may even remain in a normal state when the hole mass is large. As $U$ increases, the system enters an EPC whose character depends sensitively on the mass imbalance. Indeed, at low excitation density, the BEC regime dominates only for strongly imbalanced masses, while modest hole mobility is already sufficient to stabilize in the BCS type. This behavior specifies that itinerancy of both carriers promotes BCS pairing in the SM state. Increasing the excitation density to $n=0.1$ expands the IPC and PPC regions substantially, but only the BCS type of condensate is established. In contrast with the SC case, the BCS-BEC crossover in the SM regime is specified in EPC phase only. Taken together, these phase diagrams reveal that mass imbalance and Coulomb interaction play distinct and complementary roles in shaping the condensate landscape. The SC regime supports robust EPC and IPC BEC-BCS crossovers, whereas in the SM regime the enhanced carrier itinerancy suppresses localized pairing and confines the crossover to the EPC channel only. These results establish the mass imbalance as a key control parameter for the complex structures of PCs in microcavity systems.

\section{Conclusions}

To conclude, we have investigated the complex structure of PC states and their BCS-BEC crossover in semiconductor and semimetal microcavities in the impacts of electron-hole mass imbalance, Coulomb interaction, and excitation density. In the framework of the UHFA, we find solutions for a two-band electron-hole-photon model involving the electron-hole Coulomb attraction and light-matter coupling on an equal footing. The ground-state phase structure and luminescence spectra are then analyzed. In the SC side, a positive band gap stabilizes tightly bound excitons, leading predominantly to BEC-type condensates at low excitation density. Increasing excitation density drives a continuous crossover toward BCS-type pairing and promotes intermediate and photoniclike polaritonic condensates through enhanced the mass imbalance. The critical mass imbalance required for the crossover increases with Coulomb interaction, reflecting the robustness of excitonic binding. In contrast, the SM regime favors itinerant electron-hole pairing, and the BCS-type condensate dominates over a broad parameter range. Here, strong Coulomb interaction is necessary to establish excitoniclike coherence, while photoniclike and intermediate polaritonic states emerge and develop only if the mass imbalance is sufficiently small. The evolution of the excitonic polarization and photon luminescence spectra provides clear spectroscopic signatures of these crossover phenomena, directly linking the nature of the condensate to experimentally accessible optical responses. Our results offer a unified framework for controlling and identifying excitonic-polaritonic condensates in microcavity systems.

\section*{acknowledgement}
This research is funded by the Vietnam National Foundation for Science and Technology Development (NAFOSTED) under Grant No. 103.01-2023.43.

\section*{DATA AVAILABILITY}
The data that support the findings of this article are available from the authors upon reasonable request.

%

\begin{thebibliography}{50}%
\makeatletter
\providecommand \@ifxundefined [1]{%
 \@ifx{#1\undefined}
}%
\providecommand \@ifnum [1]{%
 \ifnum #1\expandafter \@firstoftwo
 \else \expandafter \@secondoftwo
 \fi
}%
\providecommand \@ifx [1]{%
 \ifx #1\expandafter \@firstoftwo
 \else \expandafter \@secondoftwo
 \fi
}%
\providecommand \natexlab [1]{#1}%
\providecommand \enquote  [1]{``#1''}%
\providecommand \bibnamefont  [1]{#1}%
\providecommand \bibfnamefont [1]{#1}%
\providecommand \citenamefont [1]{#1}%
\providecommand \href@noop [0]{\@secondoftwo}%
\providecommand \href [0]{\begingroup \@sanitize@url \@href}%
\providecommand \@href[1]{\@@startlink{#1}\@@href}%
\providecommand \@@href[1]{\endgroup#1\@@endlink}%
\providecommand \@sanitize@url [0]{\catcode `\\12\catcode `\$12\catcode
  `\&12\catcode `\#12\catcode `\^12\catcode `\_12\catcode `\%12\relax}%
\providecommand \@@startlink[1]{}%
\providecommand \@@endlink[0]{}%
\providecommand \url  [0]{\begingroup\@sanitize@url \@url }%
\providecommand \@url [1]{\endgroup\@href {#1}{\urlprefix }}%
\providecommand \urlprefix  [0]{URL }%
\providecommand \Eprint [0]{\href }%
\providecommand \doibase [0]{https://doi.org/}%
\providecommand \selectlanguage [0]{\@gobble}%
\providecommand \bibinfo  [0]{\@secondoftwo}%
\providecommand \bibfield  [0]{\@secondoftwo}%
\providecommand \translation [1]{[#1]}%
\providecommand \BibitemOpen [0]{}%
\providecommand \bibitemStop [0]{}%
\providecommand \bibitemNoStop [0]{.\EOS\space}%
\providecommand \EOS [0]{\spacefactor3000\relax}%
\providecommand \BibitemShut  [1]{\csname bibitem#1\endcsname}%
\let\auto@bib@innerbib\@empty
\bibitem [{\citenamefont {Mott}(1961)}]{Mo61}%
  \BibitemOpen
  \bibfield  {author} {\bibinfo {author} {\bibfnamefont {N.~F.}\ \bibnamefont
  {Mott}},\ }\bibfield  {title} {\bibinfo {title} {The transition to the
  metallic state},\ }\href@noop {} {\bibfield  {journal} {\bibinfo  {journal}
  {Philos. Mag.}\ }\textbf {\bibinfo {volume} {6}},\ \bibinfo {pages} {287}
  (\bibinfo {year} {1961})}\BibitemShut {NoStop}%
\bibitem [{\citenamefont {Knox}(1963)}]{Kno63}%
  \BibitemOpen
  \bibfield  {author} {\bibinfo {author} {\bibfnamefont {R.}~\bibnamefont
  {Knox}},\ }in\ \href@noop {} {\emph {\bibinfo {booktitle} {Solid State
  Physics}}},\ \bibinfo {editor} {edited by\ \bibinfo {editor} {\bibfnamefont
  {F.}~\bibnamefont {Seitz}}\ and\ \bibinfo {editor} {\bibfnamefont
  {D.}~\bibnamefont {Turnbull}}}\ (\bibinfo  {publisher} {Academic Press},\
  \bibinfo {address} {New York},\ \bibinfo {year} {1963})\ p.\ \bibinfo {pages}
  {Suppl. 5 p. 100}\BibitemShut {NoStop}%
\bibitem [{\citenamefont {J\'{e}rome}\ \emph {et~al.}(1967)\citenamefont
  {J\'{e}rome}, \citenamefont {Rice},\ and\ \citenamefont {Kohn}}]{JRK67}%
  \BibitemOpen
  \bibfield  {author} {\bibinfo {author} {\bibfnamefont {D.}~\bibnamefont
  {J\'{e}rome}}, \bibinfo {author} {\bibfnamefont {T.~M.}\ \bibnamefont
  {Rice}},\ and\ \bibinfo {author} {\bibfnamefont {W.}~\bibnamefont {Kohn}},\
  }\bibfield  {title} {\bibinfo {title} {Excitonic {Insulator}},\ }\href@noop
  {} {\bibfield  {journal} {\bibinfo  {journal} {Phys. Rev.}\ }\textbf
  {\bibinfo {volume} {158}},\ \bibinfo {pages} {462} (\bibinfo {year}
  {1967})}\BibitemShut {NoStop}%
\bibitem [{\citenamefont {Kohn}(1967)}]{PRL.19.439}%
  \BibitemOpen
  \bibfield  {author} {\bibinfo {author} {\bibfnamefont {W.}~\bibnamefont
  {Kohn}},\ }\bibfield  {title} {\bibinfo {title} {Excitonic phases},\
  }\href@noop {} {\bibfield  {journal} {\bibinfo  {journal} {Phys. Rev. Lett.}\
  }\textbf {\bibinfo {volume} {19}},\ \bibinfo {pages} {439} (\bibinfo {year}
  {1967})}\BibitemShut {NoStop}%
\bibitem [{\citenamefont {Halperin}\ and\ \citenamefont {Rice}(1968)}]{HR68b}%
  \BibitemOpen
  \bibfield  {author} {\bibinfo {author} {\bibfnamefont {B.~I.}\ \bibnamefont
  {Halperin}}\ and\ \bibinfo {author} {\bibfnamefont {T.~M.}\ \bibnamefont
  {Rice}},\ }\bibfield  {title} {\bibinfo {title} {Possible anomalies at a
  semimetal-semiconductor transistion},\ }\href@noop {} {\bibfield  {journal}
  {\bibinfo  {journal} {Rev. Mod. Phys.}\ }\textbf {\bibinfo {volume} {40}},\
  \bibinfo {pages} {755} (\bibinfo {year} {1968})}\BibitemShut {NoStop}%
\bibitem [{\citenamefont {Deng}\ \emph {et~al.}(2010)\citenamefont {Deng},
  \citenamefont {Haug},\ and\ \citenamefont {Yamamoto}}]{RMP.82.1489}%
  \BibitemOpen
  \bibfield  {author} {\bibinfo {author} {\bibfnamefont {H.}~\bibnamefont
  {Deng}}, \bibinfo {author} {\bibfnamefont {H.}~\bibnamefont {Haug}},\ and\
  \bibinfo {author} {\bibfnamefont {Y.}~\bibnamefont {Yamamoto}},\ }\bibfield
  {title} {\bibinfo {title} {Exciton-polariton bose-einstein condensation},\
  }\href@noop {} {\bibfield  {journal} {\bibinfo  {journal} {Rev. Mod. Phys.}\
  }\textbf {\bibinfo {volume} {82}},\ \bibinfo {pages} {1489} (\bibinfo {year}
  {2010})}\BibitemShut {NoStop}%
\bibitem [{\citenamefont {Byrnes}\ \emph {et~al.}(2014)\citenamefont {Byrnes},
  \citenamefont {Kim},\ and\ \citenamefont {Yamamoto}}]{BKY14}%
  \BibitemOpen
  \bibfield  {author} {\bibinfo {author} {\bibfnamefont {T.}~\bibnamefont
  {Byrnes}}, \bibinfo {author} {\bibfnamefont {N.~Y.}\ \bibnamefont {Kim}},\
  and\ \bibinfo {author} {\bibfnamefont {Y.}~\bibnamefont {Yamamoto}},\
  }\bibfield  {title} {\bibinfo {title} {Exciton–polariton condensates},\
  }\href@noop {} {\bibfield  {journal} {\bibinfo  {journal} {Nat. Phys.}\
  }\textbf {\bibinfo {volume} {10}},\ \bibinfo {pages} {803} (\bibinfo {year}
  {2014})}\BibitemShut {NoStop}%
\bibitem [{\citenamefont {Rahimi-Iman}(2020)}]{Iman2020}%
  \BibitemOpen
  \bibfield  {author} {\bibinfo {author} {\bibfnamefont {A.}~\bibnamefont
  {Rahimi-Iman}},\ }\href@noop {} {\emph {\bibinfo {title} {Polariton
  Physics}}}\ (\bibinfo  {publisher} {Springer},\ \bibinfo {year}
  {2020})\BibitemShut {NoStop}%
\bibitem [{\citenamefont {Christopoulos}\ \emph {et~al.}(2007)\citenamefont
  {Christopoulos}, \citenamefont {von H\"ogersthal}, \citenamefont {Grundy},
  \citenamefont {Lagoudakis}, \citenamefont {Kavokin}, \citenamefont
  {Baumberg}, \citenamefont {Christmann}, \citenamefont {Butt\'e},
  \citenamefont {Feltin}, \citenamefont {Carlin},\ and\ \citenamefont
  {Grandjean}}]{PRL.98.126405}%
  \BibitemOpen
  \bibfield  {author} {\bibinfo {author} {\bibfnamefont {S.}~\bibnamefont
  {Christopoulos}}, \bibinfo {author} {\bibfnamefont {G.~B.~H.}\ \bibnamefont
  {von H\"ogersthal}}, \bibinfo {author} {\bibfnamefont {A.~J.~D.}\
  \bibnamefont {Grundy}}, \bibinfo {author} {\bibfnamefont {P.~G.}\
  \bibnamefont {Lagoudakis}}, \bibinfo {author} {\bibfnamefont {A.~V.}\
  \bibnamefont {Kavokin}}, \bibinfo {author} {\bibfnamefont {J.~J.}\
  \bibnamefont {Baumberg}}, \bibinfo {author} {\bibfnamefont {G.}~\bibnamefont
  {Christmann}}, \bibinfo {author} {\bibfnamefont {R.}~\bibnamefont {Butt\'e}},
  \bibinfo {author} {\bibfnamefont {E.}~\bibnamefont {Feltin}}, \bibinfo
  {author} {\bibfnamefont {J.-F.}\ \bibnamefont {Carlin}},\ and\ \bibinfo
  {author} {\bibfnamefont {N.}~\bibnamefont {Grandjean}},\ }\bibfield  {title}
  {\bibinfo {title} {Room-temperature polariton lasing in semiconductor
  microcavities},\ }\href@noop {} {\bibfield  {journal} {\bibinfo  {journal}
  {Phys. Rev. Lett.}\ }\textbf {\bibinfo {volume} {98}},\ \bibinfo {pages}
  {126405} (\bibinfo {year} {2007})}\BibitemShut {NoStop}%
\bibitem [{\citenamefont {Christmann}\ \emph {et~al.}(2008)\citenamefont
  {Christmann}, \citenamefont {Butté}, \citenamefont {Feltin}, \citenamefont
  {Carlin},\ and\ \citenamefont {Grandjean}}]{APL.93.051102}%
  \BibitemOpen
  \bibfield  {author} {\bibinfo {author} {\bibfnamefont {G.}~\bibnamefont
  {Christmann}}, \bibinfo {author} {\bibfnamefont {R.}~\bibnamefont {Butté}},
  \bibinfo {author} {\bibfnamefont {E.}~\bibnamefont {Feltin}}, \bibinfo
  {author} {\bibfnamefont {J.-F.}\ \bibnamefont {Carlin}},\ and\ \bibinfo
  {author} {\bibfnamefont {N.}~\bibnamefont {Grandjean}},\ }\bibfield  {title}
  {\bibinfo {title} {Room temperature polariton lasing in a GaN/AlGaN
  multiple quantum well microcavity},\ }\href@noop {} {\bibfield  {journal}
  {\bibinfo  {journal} {Appl. Phys. Lett.}\ }\textbf {\bibinfo {volume} {93}},\
  \bibinfo {pages} {051102} (\bibinfo {year} {2008})}\BibitemShut {NoStop}%
\bibitem [{\citenamefont {K\'{e}na-Cohen}\ and\ \citenamefont
  {Forrest}(2010)}]{NPho.4.371}%
  \BibitemOpen
  \bibfield  {author} {\bibinfo {author} {\bibfnamefont {S.}~\bibnamefont
  {K\'{e}na-Cohen}}\ and\ \bibinfo {author} {\bibfnamefont {S.}~\bibnamefont
  {Forrest}},\ }\bibfield  {title} {\bibinfo {title} {Room-temperature
  polariton lasing in an organic single-crystal microcavity},\ }\href@noop {}
  {\bibfield  {journal} {\bibinfo  {journal} {Nat. Photonics}\ }\textbf
  {\bibinfo {volume} {4}},\ \bibinfo {pages} {371} (\bibinfo {year}
  {2010})}\BibitemShut {NoStop}%
\bibitem [{\citenamefont {Carusotto}\ and\ \citenamefont
  {Ciuti}(2013)}]{RMP.85.299}%
  \BibitemOpen
  \bibfield  {author} {\bibinfo {author} {\bibfnamefont {I.}~\bibnamefont
  {Carusotto}}\ and\ \bibinfo {author} {\bibfnamefont {C.}~\bibnamefont
  {Ciuti}},\ }\bibfield  {title} {\bibinfo {title} {Quantum fluids of light},\
  }\href@noop {} {\bibfield  {journal} {\bibinfo  {journal} {Rev. Mod. Phys.}\
  }\textbf {\bibinfo {volume} {85}},\ \bibinfo {pages} {299} (\bibinfo {year}
  {2013})}\BibitemShut {NoStop}%
\bibitem [{\citenamefont {Deng}\ \emph {et~al.}(2002)\citenamefont {Deng},
  \citenamefont {Weihs}, \citenamefont {Santori}, \citenamefont {Bloch},\ and\
  \citenamefont {Yamamoto}}]{DWSBY02}%
  \BibitemOpen
  \bibfield  {author} {\bibinfo {author} {\bibfnamefont {H.}~\bibnamefont
  {Deng}}, \bibinfo {author} {\bibfnamefont {G.}~\bibnamefont {Weihs}},
  \bibinfo {author} {\bibfnamefont {C.}~\bibnamefont {Santori}}, \bibinfo
  {author} {\bibfnamefont {J.}~\bibnamefont {Bloch}},\ and\ \bibinfo {author}
  {\bibfnamefont {Y.}~\bibnamefont {Yamamoto}},\ }\bibfield  {title} {\bibinfo
  {title} {Condensation of semiconductor microcavity exciton polaritons},\
  }\href@noop {} {\bibfield  {journal} {\bibinfo  {journal} {Science}\ }\textbf
  {\bibinfo {volume} {298}},\ \bibinfo {pages} {199} (\bibinfo {year}
  {2002})}\BibitemShut {NoStop}%
\bibitem [{\citenamefont {Kasprzak}\ \emph {et~al.}(2006)\citenamefont
  {Kasprzak}, \citenamefont {Richard}, \citenamefont {Kundermann},
  \citenamefont {Baas}, \citenamefont {Jeambrun}, \citenamefont {Keeling},
  \citenamefont {Marchetti}, \citenamefont {Szyma\'nska}, \citenamefont
  {Andr\'e}, \citenamefont {Staehli}, \citenamefont {Savona}, \citenamefont
  {Littlewood}, \citenamefont {Deveaud},\ and\ \citenamefont {Dang}}]{Kas06}%
  \BibitemOpen
  \bibfield  {author} {\bibinfo {author} {\bibfnamefont {J.}~\bibnamefont
  {Kasprzak}}, \bibinfo {author} {\bibfnamefont {M.}~\bibnamefont {Richard}},
  \bibinfo {author} {\bibfnamefont {S.}~\bibnamefont {Kundermann}}, \bibinfo
  {author} {\bibfnamefont {A.}~\bibnamefont {Baas}}, \bibinfo {author}
  {\bibfnamefont {P.}~\bibnamefont {Jeambrun}}, \bibinfo {author}
  {\bibfnamefont {J.~M.~J.}\ \bibnamefont {Keeling}}, \bibinfo {author}
  {\bibfnamefont {F.~M.}\ \bibnamefont {Marchetti}}, \bibinfo {author}
  {\bibfnamefont {M.~H.}\ \bibnamefont {Szyma\'nska}}, \bibinfo {author}
  {\bibfnamefont {R.}~\bibnamefont {Andr\'e}}, \bibinfo {author} {\bibfnamefont
  {J.~L.}\ \bibnamefont {Staehli}}, \bibinfo {author} {\bibfnamefont
  {V.}~\bibnamefont {Savona}}, \bibinfo {author} {\bibfnamefont {P.~B.}\
  \bibnamefont {Littlewood}}, \bibinfo {author} {\bibfnamefont
  {B.}~\bibnamefont {Deveaud}},\ and\ \bibinfo {author} {\bibfnamefont {L.~S.}\
  \bibnamefont {Dang}},\ }\bibfield  {title} {\bibinfo {title} {Bose–einstein
  condensation of exciton polaritons},\ }\href@noop {} {\bibfield  {journal}
  {\bibinfo  {journal} {Nature}\ }\textbf {\bibinfo {volume} {443}},\ \bibinfo
  {pages} {409} (\bibinfo {year} {2006})}\BibitemShut {NoStop}%
\bibitem [{\citenamefont {Hu}\ \emph {et~al.}(2021)\citenamefont {Hu},
  \citenamefont {Wang}, \citenamefont {Kim}, \citenamefont {Deng},
  \citenamefont {Brodbeck}, \citenamefont {Schneider}, \citenamefont
  {H\"ofling}, \citenamefont {Kwong},\ and\ \citenamefont
  {Binder}}]{PRX.11.011018}%
  \BibitemOpen
  \bibfield  {author} {\bibinfo {author} {\bibfnamefont {J.}~\bibnamefont
  {Hu}}, \bibinfo {author} {\bibfnamefont {Z.}~\bibnamefont {Wang}}, \bibinfo
  {author} {\bibfnamefont {S.}~\bibnamefont {Kim}}, \bibinfo {author}
  {\bibfnamefont {H.}~\bibnamefont {Deng}}, \bibinfo {author} {\bibfnamefont
  {S.}~\bibnamefont {Brodbeck}}, \bibinfo {author} {\bibfnamefont
  {C.}~\bibnamefont {Schneider}}, \bibinfo {author} {\bibfnamefont
  {S.}~\bibnamefont {H\"ofling}}, \bibinfo {author} {\bibfnamefont {N.~H.}\
  \bibnamefont {Kwong}},\ and\ \bibinfo {author} {\bibfnamefont
  {R.}~\bibnamefont {Binder}},\ }\bibfield  {title} {\bibinfo {title}
  {{Polariton Laser in the Bardeen-Cooper-Schrieffer Regime}},\ }\href@noop {}
  {\bibfield  {journal} {\bibinfo  {journal} {Phys. Rev. X}\ }\textbf {\bibinfo
  {volume} {11}},\ \bibinfo {pages} {011018} (\bibinfo {year}
  {2021})}\BibitemShut {NoStop}%
\bibitem [{\citenamefont {Bajoni}\ \emph {et~al.}(2007)\citenamefont {Bajoni},
  \citenamefont {Senellart}, \citenamefont {Lema\^{\i}tre},\ and\ \citenamefont
  {Bloch}}]{PRB.76.201305}%
  \BibitemOpen
  \bibfield  {author} {\bibinfo {author} {\bibfnamefont {D.}~\bibnamefont
  {Bajoni}}, \bibinfo {author} {\bibfnamefont {P.}~\bibnamefont {Senellart}},
  \bibinfo {author} {\bibfnamefont {A.}~\bibnamefont {Lema\^{\i}tre}},\ and\
  \bibinfo {author} {\bibfnamefont {J.}~\bibnamefont {Bloch}},\ }\bibfield
  {title} {\bibinfo {title} {Photon lasing in $\mathrm{GaAs}$ microcavity:
  Similarities with a polariton condensate},\ }\href@noop {} {\bibfield
  {journal} {\bibinfo  {journal} {Phys. Rev. B}\ }\textbf {\bibinfo {volume}
  {76}},\ \bibinfo {pages} {201305} (\bibinfo {year} {2007})}\BibitemShut
  {NoStop}%
\bibitem [{\citenamefont {Kenji}\ and\ \citenamefont {Tetsuo}(2011)}]{KO11}%
  \BibitemOpen
  \bibfield  {author} {\bibinfo {author} {\bibfnamefont {K.}~\bibnamefont
  {Kenji}}\ and\ \bibinfo {author} {\bibfnamefont {O.}~\bibnamefont {Tetsuo}},\
  }\bibfield  {title} {\bibinfo {title} {Ground-state properties of microcavity
  polariton condensates at arbitrary excitation density},\ }\href@noop {}
  {\bibfield  {journal} {\bibinfo  {journal} {Phys. Rev. B}\ }\textbf {\bibinfo
  {volume} {83}},\ \bibinfo {pages} {165319} (\bibinfo {year}
  {2011})}\BibitemShut {NoStop}%
\bibitem [{\citenamefont {Yamaguchi}\ \emph {et~al.}(2012)\citenamefont
  {Yamaguchi}, \citenamefont {Kamide}, \citenamefont {Ogawa},\ and\
  \citenamefont {Yamamoto}}]{NJP.14.065001}%
  \BibitemOpen
  \bibfield  {author} {\bibinfo {author} {\bibfnamefont {M.}~\bibnamefont
  {Yamaguchi}}, \bibinfo {author} {\bibfnamefont {K.}~\bibnamefont {Kamide}},
  \bibinfo {author} {\bibfnamefont {T.}~\bibnamefont {Ogawa}},\ and\ \bibinfo
  {author} {\bibfnamefont {Y.}~\bibnamefont {Yamamoto}},\ }\bibfield  {title}
  {\bibinfo {title} {{BEC–BCS-laser crossover in Coulomb-correlated
  electron–hole–photon systems}},\ }\href@noop {} {\bibfield  {journal}
  {\bibinfo  {journal} {New J. Phys.}\ }\textbf {\bibinfo {volume} {14}},\
  \bibinfo {pages} {065001} (\bibinfo {year} {2012})}\BibitemShut {NoStop}%
\bibitem [{\citenamefont {Yamaguchi}\ \emph {et~al.}(2015)\citenamefont
  {Yamaguchi}, \citenamefont {Nii}, \citenamefont {Kamide}, \citenamefont
  {Ogawa},\ and\ \citenamefont {Yamamoto}}]{Yamaguchi2015}%
  \BibitemOpen
  \bibfield  {author} {\bibinfo {author} {\bibfnamefont {M.}~\bibnamefont
  {Yamaguchi}}, \bibinfo {author} {\bibfnamefont {R.}~\bibnamefont {Nii}},
  \bibinfo {author} {\bibfnamefont {K.}~\bibnamefont {Kamide}}, \bibinfo
  {author} {\bibfnamefont {T.}~\bibnamefont {Ogawa}},\ and\ \bibinfo {author}
  {\bibfnamefont {Y.}~\bibnamefont {Yamamoto}},\ }\bibfield  {title} {\bibinfo
  {title} {Generating functional approach for spontaneous coherence in
  semiconductor electron-hole-photon systems},\ }\href@noop {} {\bibfield
  {journal} {\bibinfo  {journal} {Phys. Rev. B}\ }\textbf {\bibinfo {volume}
  {91}},\ \bibinfo {pages} {115129} (\bibinfo {year} {2015})}\BibitemShut
  {NoStop}%
\bibitem [{\citenamefont {Kamide}\ and\ \citenamefont
  {Ogawa}(2010)}]{kamide2010}%
  \BibitemOpen
  \bibfield  {author} {\bibinfo {author} {\bibfnamefont {K.}~\bibnamefont
  {Kamide}}\ and\ \bibinfo {author} {\bibfnamefont {T.}~\bibnamefont {Ogawa}},\
  }\bibfield  {title} {\bibinfo {title} {What determines the wave function of
  electron-hole pairs in polariton condensates},\ }\href@noop {} {\bibfield
  {journal} {\bibinfo  {journal} {Phys. Rev. Lett.}\ }\textbf {\bibinfo
  {volume} {105}},\ \bibinfo {pages} {056401} (\bibinfo {year}
  {2010})}\BibitemShut {NoStop}%
\bibitem [{\citenamefont {Phan}\ \emph {et~al.}(2016)\citenamefont {Phan},
  \citenamefont {Becker},\ and\ \citenamefont {Fehske}}]{PBF16}%
  \BibitemOpen
  \bibfield  {author} {\bibinfo {author} {\bibfnamefont {V.-N.}\ \bibnamefont
  {Phan}}, \bibinfo {author} {\bibfnamefont {K.~W.}\ \bibnamefont {Becker}},\
  and\ \bibinfo {author} {\bibfnamefont {H.}~\bibnamefont {Fehske}},\
  }\bibfield  {title} {\bibinfo {title} {Ground-state and spectral signatures
  of cavity exciton-polariton condensates},\ }\href@noop {} {\bibfield
  {journal} {\bibinfo  {journal} {Phys. Rev. B}\ }\textbf {\bibinfo {volume}
  {93}},\ \bibinfo {pages} {075138} (\bibinfo {year} {2016})}\BibitemShut
  {NoStop}%
\bibitem [{\citenamefont {Ninh}\ and\ \citenamefont
  {Phan}(2019)}]{PhysB.573.72}%
  \BibitemOpen
  \bibfield  {author} {\bibinfo {author} {\bibfnamefont {Q.-H.}\ \bibnamefont
  {Ninh}}\ and\ \bibinfo {author} {\bibfnamefont {V.-N.}\ \bibnamefont
  {Phan}},\ }\bibfield  {title} {\bibinfo {title} {{BCS-BEC crossovers of
  microcavity exciton-polariton condensates}},\ }\href@noop {} {\bibfield
  {journal} {\bibinfo  {journal} {Physica B: Condense Matt.}\ }\textbf
  {\bibinfo {volume} {573}},\ \bibinfo {pages} {72} (\bibinfo {year}
  {2019})}\BibitemShut {NoStop}%
\bibitem [{\citenamefont {Ejima}\ \emph {et~al.}(2014)\citenamefont {Ejima},
  \citenamefont {Kaneko}, \citenamefont {Ohta},\ and\ \citenamefont
  {Fehske}}]{EKOF14}%
  \BibitemOpen
  \bibfield  {author} {\bibinfo {author} {\bibfnamefont {S.}~\bibnamefont
  {Ejima}}, \bibinfo {author} {\bibfnamefont {T.}~\bibnamefont {Kaneko}},
  \bibinfo {author} {\bibfnamefont {T.}~\bibnamefont {Ohta}},\ and\ \bibinfo
  {author} {\bibfnamefont {H.}~\bibnamefont {Fehske}},\ }\bibfield  {title}
  {\bibinfo {title} {{Order, Criticality, and Excitations in the Extended
  Falicov-Kimball Model}},\ }\href@noop {} {\bibfield  {journal} {\bibinfo
  {journal} {Phys. Rev. Lett.}\ }\textbf {\bibinfo {volume} {112}},\ \bibinfo
  {pages} {026401} (\bibinfo {year} {2014})}\BibitemShut {NoStop}%
\bibitem [{\citenamefont {Do}\ \emph {et~al.}(2024)\citenamefont {Do},
  \citenamefont {Nguyen},\ and\ \citenamefont {Phan}}]{PRB.109.085105}%
  \BibitemOpen
  \bibfield  {author} {\bibinfo {author} {\bibfnamefont {T.-H.-H.}\
  \bibnamefont {Do}}, \bibinfo {author} {\bibfnamefont {T.-H.}\ \bibnamefont
  {Nguyen}},\ and\ \bibinfo {author} {\bibfnamefont {V.-N.}\ \bibnamefont
  {Phan}},\ }\bibfield  {title} {\bibinfo {title} {Mass imbalance and
  electron-phonon correlation impact on the excitonic insulator state in
  semimetal/semiconducting materials},\ }\href@noop {} {\bibfield  {journal}
  {\bibinfo  {journal} {Phys. Rev. B}\ }\textbf {\bibinfo {volume} {109}},\
  \bibinfo {pages} {085105} (\bibinfo {year} {2024})}\BibitemShut {NoStop}%
\bibitem [{\citenamefont {Do}\ and\ \citenamefont
  {Phan}(2024)}]{PRB.110.235143}%
  \BibitemOpen
  \bibfield  {author} {\bibinfo {author} {\bibfnamefont {T.-H.-H.}\
  \bibnamefont {Do}}\ and\ \bibinfo {author} {\bibfnamefont {V.-N.}\
  \bibnamefont {Phan}},\ }\bibfield  {title} {\bibinfo {title} {Thermal
  fluctuations and mass imbalance in low-energy excitonic condensate
  excitations},\ }\href@noop {} {\bibfield  {journal} {\bibinfo  {journal}
  {Phys. Rev. B}\ }\textbf {\bibinfo {volume} {110}},\ \bibinfo {pages}
  {235143} (\bibinfo {year} {2024})}\BibitemShut {NoStop}%
\bibitem [{\citenamefont {Nguyen}\ \emph {et~al.}(2025)\citenamefont {Nguyen},
  \citenamefont {Do},\ and\ \citenamefont {Phan}}]{PRB.111.245111}%
  \BibitemOpen
  \bibfield  {author} {\bibinfo {author} {\bibfnamefont {T.-H.}\ \bibnamefont
  {Nguyen}}, \bibinfo {author} {\bibfnamefont {T.-H.-H.}\ \bibnamefont {Do}},\
  and\ \bibinfo {author} {\bibfnamefont {V.-N.}\ \bibnamefont {Phan}},\
  }\bibfield  {title} {\bibinfo {title} {Quantum coherent states of
  mass-imbalanced electron-hole system within optical microcavities},\
  }\href@noop {} {\bibfield  {journal} {\bibinfo  {journal} {Phys. Rev. B}\
  }\textbf {\bibinfo {volume} {111}},\ \bibinfo {pages} {245111} (\bibinfo
  {year} {2025})}\BibitemShut {NoStop}%
\bibitem [{\citenamefont {Kormányos}\ \emph {et~al.}(2015)\citenamefont
  {Kormányos}, \citenamefont {Burkard}, \citenamefont {Gmitra}, \citenamefont
  {Fabian}, \citenamefont {Zólyomi}, \citenamefont {Drummond},\ and\
  \citenamefont {Fal’ko}}]{2DM.2.022001}%
  \BibitemOpen
  \bibfield  {author} {\bibinfo {author} {\bibfnamefont {A.}~\bibnamefont
  {Kormányos}}, \bibinfo {author} {\bibfnamefont {G.}~\bibnamefont {Burkard}},
  \bibinfo {author} {\bibfnamefont {M.}~\bibnamefont {Gmitra}}, \bibinfo
  {author} {\bibfnamefont {J.}~\bibnamefont {Fabian}}, \bibinfo {author}
  {\bibfnamefont {V.}~\bibnamefont {Zólyomi}}, \bibinfo {author}
  {\bibfnamefont {N.~D.}\ \bibnamefont {Drummond}},\ and\ \bibinfo {author}
  {\bibfnamefont {V.}~\bibnamefont {Fal’ko}},\ }\bibfield  {title} {\bibinfo
  {title} {k·p theory for two-dimensional transition metal dichalcogenide
  semiconductors},\ }\href@noop {} {\bibfield  {journal} {\bibinfo  {journal}
  {2D Materials}\ }\textbf {\bibinfo {volume} {2}},\ \bibinfo {pages} {022001}
  (\bibinfo {year} {2015})}\BibitemShut {NoStop}%
\bibitem [{\citenamefont {Chang}\ and\ \citenamefont
  {Banerjee}(2014)}]{Jiwon2014}%
  \BibitemOpen
  \bibfield  {author} {\bibinfo {author} {\bibfnamefont {J.}~\bibnamefont
  {Chang}}\ and\ \bibinfo {author} {\bibfnamefont {S.}~\bibnamefont
  {Banerjee}},\ }\bibfield  {title} {\bibinfo {title} {{Ballistic performance
  comparison of monolayer transition metal dichalcogenide MX2 (M = Mo, W; X =
  S, Se, Te) metal-oxide-semiconductor field effect transistors}},\ }\href@noop
  {} {\bibfield  {journal} {\bibinfo  {journal} {J. Appl. Phys.}\ }\textbf
  {\bibinfo {volume} {115}},\ \bibinfo {pages} {084506} (\bibinfo {year}
  {2014})}\BibitemShut {NoStop}%
\bibitem [{\citenamefont {Nakwaski}(1995)}]{PhysicaB.210.1}%
  \BibitemOpen
  \bibfield  {author} {\bibinfo {author} {\bibfnamefont {W.}~\bibnamefont
  {Nakwaski}},\ }\bibfield  {title} {\bibinfo {title} {{Effective masses of
  electrons and heavy holes in GaAs, InAs, A1As and their ternary compounds}},\
  }\href@noop {} {\bibfield  {journal} {\bibinfo  {journal} {Physica B}\
  }\textbf {\bibinfo {volume} {210}},\ \bibinfo {pages} {1} (\bibinfo {year}
  {1995})}\BibitemShut {NoStop}%
\bibitem [{\citenamefont {Pau}\ \emph {et~al.}(1995)\citenamefont {Pau},
  \citenamefont {Bjork}, \citenamefont {Jacobson}, \citenamefont {Cao},\ and\
  \citenamefont {Yamamoto}}]{SGJ1995}%
  \BibitemOpen
  \bibfield  {author} {\bibinfo {author} {\bibfnamefont {S.}~\bibnamefont
  {Pau}}, \bibinfo {author} {\bibfnamefont {G.}~\bibnamefont {Bjork}}, \bibinfo
  {author} {\bibfnamefont {J.}~\bibnamefont {Jacobson}}, \bibinfo {author}
  {\bibfnamefont {H.}~\bibnamefont {Cao}},\ and\ \bibinfo {author}
  {\bibfnamefont {Y.}~\bibnamefont {Yamamoto}},\ }\bibfield  {title} {\bibinfo
  {title} {Simulated emission of a microcavity dressed exciton and suppression
  of phonon scattering},\ }\href@noop {} {\bibfield  {journal} {\bibinfo
  {journal} {Phys. Rev. B}\ }\textbf {\bibinfo {volume} {51}},\ \bibinfo
  {pages} {7090} (\bibinfo {year} {1995})}\BibitemShut {NoStop}%
\bibitem [{\citenamefont {Suzuki}\ and\ \citenamefont
  {Uenoyama}(1995)}]{Suzuki1995}%
  \BibitemOpen
  \bibfield  {author} {\bibinfo {author} {\bibfnamefont {M.}~\bibnamefont
  {Suzuki}}\ and\ \bibinfo {author} {\bibfnamefont {T.}~\bibnamefont
  {Uenoyama}},\ }\bibfield  {title} {\bibinfo {title} {First-principles
  calculations of effective-mass parameters of a1n and gan},\ }\href@noop {}
  {\bibfield  {journal} {\bibinfo  {journal} {Phys. Rev. B}\ }\textbf {\bibinfo
  {volume} {52}},\ \bibinfo {pages} {8132} (\bibinfo {year}
  {1995})}\BibitemShut {NoStop}%
\bibitem [{\citenamefont {Bronold}\ and\ \citenamefont {Fehske}(2006)}]{BF06}%
  \BibitemOpen
  \bibfield  {author} {\bibinfo {author} {\bibfnamefont {F.~X.}\ \bibnamefont
  {Bronold}}\ and\ \bibinfo {author} {\bibfnamefont {H.}~\bibnamefont
  {Fehske}},\ }\bibfield  {title} {\bibinfo {title} {Possibility of an
  excitonic insulator at the semiconductor-semimetal transition},\ }\href@noop
  {} {\bibfield  {journal} {\bibinfo  {journal} {Phys. Rev. B}\ }\textbf
  {\bibinfo {volume} {74}},\ \bibinfo {pages} {165107} (\bibinfo {year}
  {2006})}\BibitemShut {NoStop}%
\bibitem [{\citenamefont {Kaneko}\ \emph {et~al.}(2013)\citenamefont {Kaneko},
  \citenamefont {Ejima}, \citenamefont {Fehske},\ and\ \citenamefont
  {Ohta}}]{KEFO13}%
  \BibitemOpen
  \bibfield  {author} {\bibinfo {author} {\bibfnamefont {T.}~\bibnamefont
  {Kaneko}}, \bibinfo {author} {\bibfnamefont {S.}~\bibnamefont {Ejima}},
  \bibinfo {author} {\bibfnamefont {H.}~\bibnamefont {Fehske}},\ and\ \bibinfo
  {author} {\bibfnamefont {Y.}~\bibnamefont {Ohta}},\ }\bibfield  {title}
  {\bibinfo {title} {Exact-diagonalization study of exciton condensation in
  electron bilayers},\ }\href@noop {} {\bibfield  {journal} {\bibinfo
  {journal} {Phys. Rev. B}\ }\textbf {\bibinfo {volume} {88}},\ \bibinfo
  {pages} {035312} (\bibinfo {year} {2013})}\BibitemShut {NoStop}%
\bibitem [{\citenamefont {Li}\ \emph {et~al.}(2021)\citenamefont {Li},
  \citenamefont {Parish},\ and\ \citenamefont {Levinsen}}]{PRB.104.245404}%
  \BibitemOpen
  \bibfield  {author} {\bibinfo {author} {\bibfnamefont {G.}~\bibnamefont
  {Li}}, \bibinfo {author} {\bibfnamefont {M.~M.}\ \bibnamefont {Parish}},\
  and\ \bibinfo {author} {\bibfnamefont {J.}~\bibnamefont {Levinsen}},\
  }\bibfield  {title} {\bibinfo {title} {Microscopic calculation of polariton
  scattering in semiconductor microcavities},\ }\href@noop {} {\bibfield
  {journal} {\bibinfo  {journal} {Phys. Rev. B}\ }\textbf {\bibinfo {volume}
  {104}},\ \bibinfo {pages} {245404} (\bibinfo {year} {2021})}\BibitemShut
  {NoStop}%
\bibitem [{\citenamefont {Tiene}\ \emph {et~al.}(2020)\citenamefont {Tiene},
  \citenamefont {Levinsen}, \citenamefont {Parish}, \citenamefont {MacDonald},
  \citenamefont {Keeling},\ and\ \citenamefont {Marchetti}}]{PRR.2.023089}%
  \BibitemOpen
  \bibfield  {author} {\bibinfo {author} {\bibfnamefont {A.}~\bibnamefont
  {Tiene}}, \bibinfo {author} {\bibfnamefont {J.}~\bibnamefont {Levinsen}},
  \bibinfo {author} {\bibfnamefont {M.~M.}\ \bibnamefont {Parish}}, \bibinfo
  {author} {\bibfnamefont {A.~H.}\ \bibnamefont {MacDonald}}, \bibinfo {author}
  {\bibfnamefont {J.}~\bibnamefont {Keeling}},\ and\ \bibinfo {author}
  {\bibfnamefont {F.~M.}\ \bibnamefont {Marchetti}},\ }\bibfield  {title}
  {\bibinfo {title} {Extremely imbalanced two-dimensional electron-hole-photon
  systems},\ }\href@noop {} {\bibfield  {journal} {\bibinfo  {journal} {Phys.
  Rev. Res.}\ }\textbf {\bibinfo {volume} {2}},\ \bibinfo {pages} {023089}
  (\bibinfo {year} {2020})}\BibitemShut {NoStop}%
\bibitem [{\citenamefont {Tiene}(2023)}]{PhDTh-Tiene}%
  \BibitemOpen
  \bibfield  {author} {\bibinfo {author} {\bibfnamefont {A.}~\bibnamefont
  {Tiene}},\ }\emph {\bibinfo {title} {{Charged Polaritons in Two-Dimensional
  Semiconductors}}},\ \href@noop {} {Ph.D. thesis},\ \bibinfo  {school}
  {Universidad Autónoma de Madrid} (\bibinfo {year} {2023})\BibitemShut
  {NoStop}%
\bibitem [{\citenamefont {Schneider}\ and\ \citenamefont
  {Czycholl}(2008)}]{SC08}%
  \BibitemOpen
  \bibfield  {author} {\bibinfo {author} {\bibfnamefont {C.}~\bibnamefont
  {Schneider}}\ and\ \bibinfo {author} {\bibfnamefont {G.}~\bibnamefont
  {Czycholl}},\ }\href@noop {} {\bibfield  {journal} {\bibinfo  {journal} {Eur.
  Phys. J. B}\ }\textbf {\bibinfo {volume} {64}},\ \bibinfo {pages} {43}
  (\bibinfo {year} {2008})}\BibitemShut {NoStop}%
\bibitem [{\citenamefont {Do}\ and\ \citenamefont
  {Phan}(2023)}]{PRB.107.115106}%
  \BibitemOpen
  \bibfield  {author} {\bibinfo {author} {\bibfnamefont {T.-H.-H.}\
  \bibnamefont {Do}}\ and\ \bibinfo {author} {\bibfnamefont {V.-N.}\
  \bibnamefont {Phan}},\ }\bibfield  {title} {\bibinfo {title} {Electron-phonon
  correlations inducing excitonic excitations in semimetal and semiconducting
  materials},\ }\href@noop {} {\bibfield  {journal} {\bibinfo  {journal} {Phys.
  Rev. B}\ }\textbf {\bibinfo {volume} {107}},\ \bibinfo {pages} {115106}
  (\bibinfo {year} {2023})}\BibitemShut {NoStop}%
\bibitem [{\citenamefont {Shi}\ \emph {et~al.}(1994)\citenamefont {Shi},
  \citenamefont {Verechaka},\ and\ \citenamefont {Griffin}}]{PRB.50.1119}%
  \BibitemOpen
  \bibfield  {author} {\bibinfo {author} {\bibfnamefont {H.}~\bibnamefont
  {Shi}}, \bibinfo {author} {\bibfnamefont {G.}~\bibnamefont {Verechaka}},\
  and\ \bibinfo {author} {\bibfnamefont {A.}~\bibnamefont {Griffin}},\
  }\bibfield  {title} {\bibinfo {title} {{Theory of the decay luminescence
  spectrum of a Bose-condensed interacting exciton gas}},\ }\href@noop {}
  {\bibfield  {journal} {\bibinfo  {journal} {Phys. Rev. B}\ }\textbf {\bibinfo
  {volume} {50}},\ \bibinfo {pages} {1119} (\bibinfo {year}
  {1994})}\BibitemShut {NoStop}%
\bibitem [{\citenamefont {Stolz}\ and\ \citenamefont
  {Semkat}(2010)}]{PRB.81.081302}%
  \BibitemOpen
  \bibfield  {author} {\bibinfo {author} {\bibfnamefont {H.}~\bibnamefont
  {Stolz}}\ and\ \bibinfo {author} {\bibfnamefont {D.}~\bibnamefont {Semkat}},\
  }\bibfield  {title} {\bibinfo {title} {{Unique signatures for Bose-Einstein
  condensation in the decay luminescence lineshape of weakly interacting
  excitons in a potential trap}},\ }\href@noop {} {\bibfield  {journal}
  {\bibinfo  {journal} {Phys. Rev. B}\ }\textbf {\bibinfo {volume} {81}},\
  \bibinfo {pages} {081302} (\bibinfo {year} {2010})}\BibitemShut {NoStop}%
\bibitem [{\citenamefont {Stolz}\ \emph {et~al.}(2012)\citenamefont {Stolz},
  \citenamefont {Schwartz}, \citenamefont {Kieseling}, \citenamefont {Som},
  \citenamefont {Kaupsch}, \citenamefont {Sobkowiak}, \citenamefont {Semkat},
  \citenamefont {Naka}, \citenamefont {Koch},\ and\ \citenamefont
  {Fehske}}]{SSKSKSSNKF12}%
  \BibitemOpen
  \bibfield  {author} {\bibinfo {author} {\bibfnamefont {H.}~\bibnamefont
  {Stolz}}, \bibinfo {author} {\bibfnamefont {R.}~\bibnamefont {Schwartz}},
  \bibinfo {author} {\bibfnamefont {F.}~\bibnamefont {Kieseling}}, \bibinfo
  {author} {\bibfnamefont {S.}~\bibnamefont {Som}}, \bibinfo {author}
  {\bibfnamefont {M.}~\bibnamefont {Kaupsch}}, \bibinfo {author} {\bibfnamefont
  {S.}~\bibnamefont {Sobkowiak}}, \bibinfo {author} {\bibfnamefont
  {D.}~\bibnamefont {Semkat}}, \bibinfo {author} {\bibfnamefont
  {N.}~\bibnamefont {Naka}}, \bibinfo {author} {\bibfnamefont {T.}~\bibnamefont
  {Koch}},\ and\ \bibinfo {author} {\bibfnamefont {H.}~\bibnamefont {Fehske}},\
  }\bibfield  {title} {\bibinfo {title} {Condensation of excitons in {Cu$_2$O}
  at ultracold temperatures: experiment and theory},\ }\href@noop {} {\bibfield
   {journal} {\bibinfo  {journal} {New J. Phys.}\ }\textbf {\bibinfo {volume}
  {14}},\ \bibinfo {pages} {105007} (\bibinfo {year} {2012})}\BibitemShut
  {NoStop}%
\bibitem [{\citenamefont {Bui}\ and\ \citenamefont {Phan}(2017)}]{NHAM2017}%
  \BibitemOpen
  \bibfield  {author} {\bibinfo {author} {\bibfnamefont {D.}~\bibnamefont
  {Bui}}\ and\ \bibinfo {author} {\bibfnamefont {V.}~\bibnamefont {Phan}},\
  }\bibfield  {title} {\bibinfo {title} {Thermal fluctuations in a system of
  microcavity exciton-polariton condensations},\ }\href@noop {} {\bibfield
  {journal} {\bibinfo  {journal} {Phys. Status Solidi B}\ }\textbf {\bibinfo
  {volume} {254}},\ \bibinfo {pages} {1600359} (\bibinfo {year}
  {2017})}\BibitemShut {NoStop}%
\bibitem [{\citenamefont {Phan}\ \emph {et~al.}(2010)\citenamefont {Phan},
  \citenamefont {Becker},\ and\ \citenamefont {Fehske}}]{PBF10}%
  \BibitemOpen
  \bibfield  {author} {\bibinfo {author} {\bibfnamefont {N.~V.}\ \bibnamefont
  {Phan}}, \bibinfo {author} {\bibfnamefont {K.~W.}\ \bibnamefont {Becker}},\
  and\ \bibinfo {author} {\bibfnamefont {H.}~\bibnamefont {Fehske}},\
  }\bibfield  {title} {\bibinfo {title} {Spectral signatures of the {BCS-BEC}
  crossover in the excitonic insulator phase of the extended {Falicov-Kimball}
  model},\ }\href@noop {} {\bibfield  {journal} {\bibinfo  {journal} {Phys.
  Rev. B}\ }\textbf {\bibinfo {volume} {81}},\ \bibinfo {pages} {205117}
  (\bibinfo {year} {2010})}\BibitemShut {NoStop}%
\bibitem [{\citenamefont {Ihle}\ \emph {et~al.}(2008)\citenamefont {Ihle},
  \citenamefont {Pfafferott}, \citenamefont {Burovski}, \citenamefont
  {Bronold},\ and\ \citenamefont {Fehske}}]{IPBBF08}%
  \BibitemOpen
  \bibfield  {author} {\bibinfo {author} {\bibfnamefont {D.}~\bibnamefont
  {Ihle}}, \bibinfo {author} {\bibfnamefont {M.}~\bibnamefont {Pfafferott}},
  \bibinfo {author} {\bibfnamefont {E.}~\bibnamefont {Burovski}}, \bibinfo
  {author} {\bibfnamefont {F.~X.}\ \bibnamefont {Bronold}},\ and\ \bibinfo
  {author} {\bibfnamefont {H.}~\bibnamefont {Fehske}},\ }\bibfield  {title}
  {\bibinfo {title} {Bound state formation and nature of the excitonic
  insulator phase in the extended {Falicov-Kimball} model},\ }\href@noop {}
  {\bibfield  {journal} {\bibinfo  {journal} {Phys. Rev. B}\ }\textbf {\bibinfo
  {volume} {78}},\ \bibinfo {pages} {193103} (\bibinfo {year}
  {2008})}\BibitemShut {NoStop}%
\bibitem [{\citenamefont {Seki}\ \emph {et~al.}(2011)\citenamefont {Seki},
  \citenamefont {Eder},\ and\ \citenamefont {Ohta}}]{SEO11}%
  \BibitemOpen
  \bibfield  {author} {\bibinfo {author} {\bibfnamefont {K.}~\bibnamefont
  {Seki}}, \bibinfo {author} {\bibfnamefont {R.}~\bibnamefont {Eder}},\ and\
  \bibinfo {author} {\bibfnamefont {Y.}~\bibnamefont {Ohta}},\ }\bibfield
  {title} {\bibinfo {title} {{BCS-BEC} crossover in the extended
  {Falicov-Kimball} model: Variational cluster approach},\ }\href@noop {}
  {\bibfield  {journal} {\bibinfo  {journal} {Phys. Rev. B}\ }\textbf {\bibinfo
  {volume} {84}},\ \bibinfo {pages} {245106} (\bibinfo {year}
  {2011})}\BibitemShut {NoStop}%
\bibitem [{\citenamefont {Zenker}\ \emph {et~al.}(2012)\citenamefont {Zenker},
  \citenamefont {Ihle}, \citenamefont {Bronold},\ and\ \citenamefont
  {Fehske}}]{ZIBF12}%
  \BibitemOpen
  \bibfield  {author} {\bibinfo {author} {\bibfnamefont {B.}~\bibnamefont
  {Zenker}}, \bibinfo {author} {\bibfnamefont {D.}~\bibnamefont {Ihle}},
  \bibinfo {author} {\bibfnamefont {F.~X.}\ \bibnamefont {Bronold}},\ and\
  \bibinfo {author} {\bibfnamefont {H.}~\bibnamefont {Fehske}},\ }\bibfield
  {title} {\bibinfo {title} {Electron-hole pair condensation at the
  semimetal-semiconductor transition: A {BCS-BEC} crossover scenario},\
  }\href@noop {} {\bibfield  {journal} {\bibinfo  {journal} {Phys. Rev. B}\
  }\textbf {\bibinfo {volume} {85}},\ \bibinfo {pages} {121102(R)} (\bibinfo
  {year} {2012})}\BibitemShut {NoStop}%
\bibitem [{\citenamefont {Dicke}(1954)}]{dick1954}%
  \BibitemOpen
  \bibfield  {author} {\bibinfo {author} {\bibfnamefont {R.~H.}\ \bibnamefont
  {Dicke}},\ }\bibfield  {title} {\bibinfo {title} {Coherence in spontaneous
  radiation processes},\ }\href@noop {} {\bibfield  {journal} {\bibinfo
  {journal} {Phys. Rev.}\ }\textbf {\bibinfo {volume} {93}},\ \bibinfo {pages}
  {99} (\bibinfo {year} {1954})}\BibitemShut {NoStop}%
\bibitem [{\citenamefont {Eastham}\ and\ \citenamefont
  {Littlewood}(2001)}]{PRB.64.235101}%
  \BibitemOpen
  \bibfield  {author} {\bibinfo {author} {\bibfnamefont {P.~R.}\ \bibnamefont
  {Eastham}}\ and\ \bibinfo {author} {\bibfnamefont {P.~B.}\ \bibnamefont
  {Littlewood}},\ }\bibfield  {title} {\bibinfo {title} {Bose condensation of
  cavity polaritons beyond the linear regime: The thermal equilibrium of a
  model microcavity},\ }\href@noop {} {\bibfield  {journal} {\bibinfo
  {journal} {Phys. Rev. B}\ }\textbf {\bibinfo {volume} {64}},\ \bibinfo
  {pages} {235101} (\bibinfo {year} {2001})}\BibitemShut {NoStop}%
\bibitem [{\citenamefont {Szyma\ifmmode~\acute{n}\else \'{n}\fi{}ska}\ \emph
  {et~al.}(2006)\citenamefont {Szyma\ifmmode~\acute{n}\else \'{n}\fi{}ska},
  \citenamefont {Keeling},\ and\ \citenamefont {Littlewood}}]{PRL.96.230602}%
  \BibitemOpen
  \bibfield  {author} {\bibinfo {author} {\bibfnamefont {M.~H.}\ \bibnamefont
  {Szyma\ifmmode~\acute{n}\else \'{n}\fi{}ska}}, \bibinfo {author}
  {\bibfnamefont {J.}~\bibnamefont {Keeling}},\ and\ \bibinfo {author}
  {\bibfnamefont {P.~B.}\ \bibnamefont {Littlewood}},\ }\bibfield  {title}
  {\bibinfo {title} {Nonequilibrium quantum condensation in an incoherently
  pumped dissipative system},\ }\href@noop {} {\bibfield  {journal} {\bibinfo
  {journal} {Phys. Rev. Lett.}\ }\textbf {\bibinfo {volume} {96}},\ \bibinfo
  {pages} {230602} (\bibinfo {year} {2006})}\BibitemShut {NoStop}%
\bibitem [{\citenamefont {Bui}\ and\ \citenamefont {Phan}(2016)}]{NHAM2016}%
  \BibitemOpen
  \bibfield  {author} {\bibinfo {author} {\bibfnamefont {D.}~\bibnamefont
  {Bui}}\ and\ \bibinfo {author} {\bibfnamefont {V.}~\bibnamefont {Phan}},\
  }\bibfield  {title} {\bibinfo {title} {Phase diagram of microcavity
  exciton-polariton condensates},\ }\href@noop {} {\bibfield  {journal}
  {\bibinfo  {journal} {Europhys. Lett.}\ }\textbf {\bibinfo {volume} {116}},\
  \bibinfo {pages} {57004} (\bibinfo {year} {2016})}\BibitemShut {NoStop}%
\end{thebibliography}

\end{document}